\documentclass[a4paper,11pt]{article}
\usepackage{jcappub} 

\title{\boldmath Topology-based fast neutron background suppression and sterile-neutrino sensitivity in a segmented reactor ${\overline{\ensuremath{\nu}}}_{e}$ detector}

\author{Roni Dey}
\affiliation{State Key Laboratory of Dark Matter Physics, Key Laboratory for Particle Astrophysics and
Cosmology (MoE), Shanghai Key Laboratory for Particle Physics and Cosmology, New Cornerstone
Science Laboratory, Tsung-Dao Lee Institute, Shanghai Jiao Tong University, Shanghai 201210, China}
\emailAdd{dmphyroni@gmail.com}

\abstract{Reactor electron antineutrinos (${\overline{\ensuremath{\nu}}}_{e}$) are commonly detected through the inverse beta decay (IBD) interaction, in which a prompt positron signal is followed by a delayed neutron capture signal. For segmented organic scintillator detectors operated at shallow depth or above ground, correlated fast neutron backgrounds constitute one of the dominant sources of background because proton recoils can closely mimic the prompt positron signature of IBD events. In this work, we investigate a topology-based machine learning selection using the Indian Scintillator Matrix for Reactor Anti-Neutrinos (ISMRAN) detector as a case study. A multilayer perceptron (MLP) neural network is trained using reconstructed topological observables obtained from detailed \textsc{Geant4} simulations. The classifier operating point is determined from an independent validation sample using the $\mathrm{F_{1}}$ score and bootstrap resampling, yielding a fixed threshold of $\mathrm{P_{IBD}>}0.40$. Applied without further optimization to an independent testing sample, the MLP retains $\mathrm{(83.4\pm 0.7)\%}$ of IBD events while rejecting $\mathrm{(77.3\pm 0.7)\%}$ of fast neutron events. In a controlled spectral benchmark using an independent AmBe fast neutron spectrum, different from the uniform ($2$-$12~\mathrm{MeV}$) distribution used for training, the selection reduces the net candidate-spectrum deformation by $\mathrm{84\%}$. The energy dependent signal acceptance and residual fast neutron component are subsequently incorporated into a binned likelihood, which recovers the injected IBD signal strength as $\mathrm{\hat{\alpha}=0.993^{+0.053}_{-0.048}}$ with an empirical goodness-of-fit probability of $0.434$. The expected sterile-neutrino sensitivity for the ISMRAN detector is evaluated at baselines of 13 m and 18 m, with two years of exposure at each baseline. For the nominal pre-MLP $S/B=1:2$ benchmark and the background model, retaining the classifier score in a joint ($\mathrm{E_{prompt}\times P_{IBD}}$) likelihood improves the 95\% C.L. sensitivity to $\sin^{2}2\theta_{14}$ by a logarithmically averaged 7.9\% over $0.1\leq\Delta m_{41}^{2}\leq1~\mathrm{eV}^{2}$ relative to an energy-only likelihood with the same fast neutron contamination. These results indicate that topology information is particularly valuable where oscillation-induced spectral modifications compete with correlated fast neutron backgrounds.}

\begin{document}
\maketitle
\flushbottom

\section{Introduction}
\label{intro}

Reactor antineutrinos (${\overline{\ensuremath{\nu}}}_{e}$) provide a unique probe of weak interactions and neutrino properties at MeV energies. Their intense and well characterized flux has enabled precision measurements of neutrino oscillation parameters and investigations for new physics, including the Reactor Antineutrino Anomaly (RAA) and light sterile neutrinos~\cite{STEREO2023Final}. Beyond fundamental physics, reactor ${\overline{\ensuremath{\nu}}}_{e}$ detection offers a non-intrusive technique for reactor monitoring and nuclear safeguards. Reactor ${\overline{\ensuremath{\nu}}}_{e}$ are predominantly detected through the inverse beta decay (IBD) process (charged-current interaction), which produces a characteristic prompt positron signal followed by a delayed neutron capture signal. This distinctive time correlated signature efficiently suppresses accidental backgrounds and has been successfully employed in reactor ${\overline{\ensuremath{\nu}}}_{e}$ experiments such as KamLAND, Daya Bay, Double Chooz, and RENO~\cite{kamland,dayabay,doublechooz,reno}. However, correlated backgrounds remain a major challenge, particularly for detectors operating at shallow depth or above ground. Among these, fast neutrons are particularly problematic because they can reproduce both components of the IBD prompt-delayed signature.

Segmented organic scintillator detectors provide a natural handle for such discrimination. These detectors have emerged as a powerful technology for very short baseline reactor ${\overline{\ensuremath{\nu}}}_{e}$ experiments. Many experiments such as PROSPECT~\cite{prospect}, STEREO~\cite{stereo}, and DANSS~\cite{danss} exploit fine detector segmentation to reconstruct event topology and improve background rejection through event topology reconstruction. The high granularity of segmented detectors offers access to detailed spatial and energy deposition information that can be exploited to further improve signal identification.

\begin{figure}
	\centering 
	\includegraphics[width=0.42\textwidth]{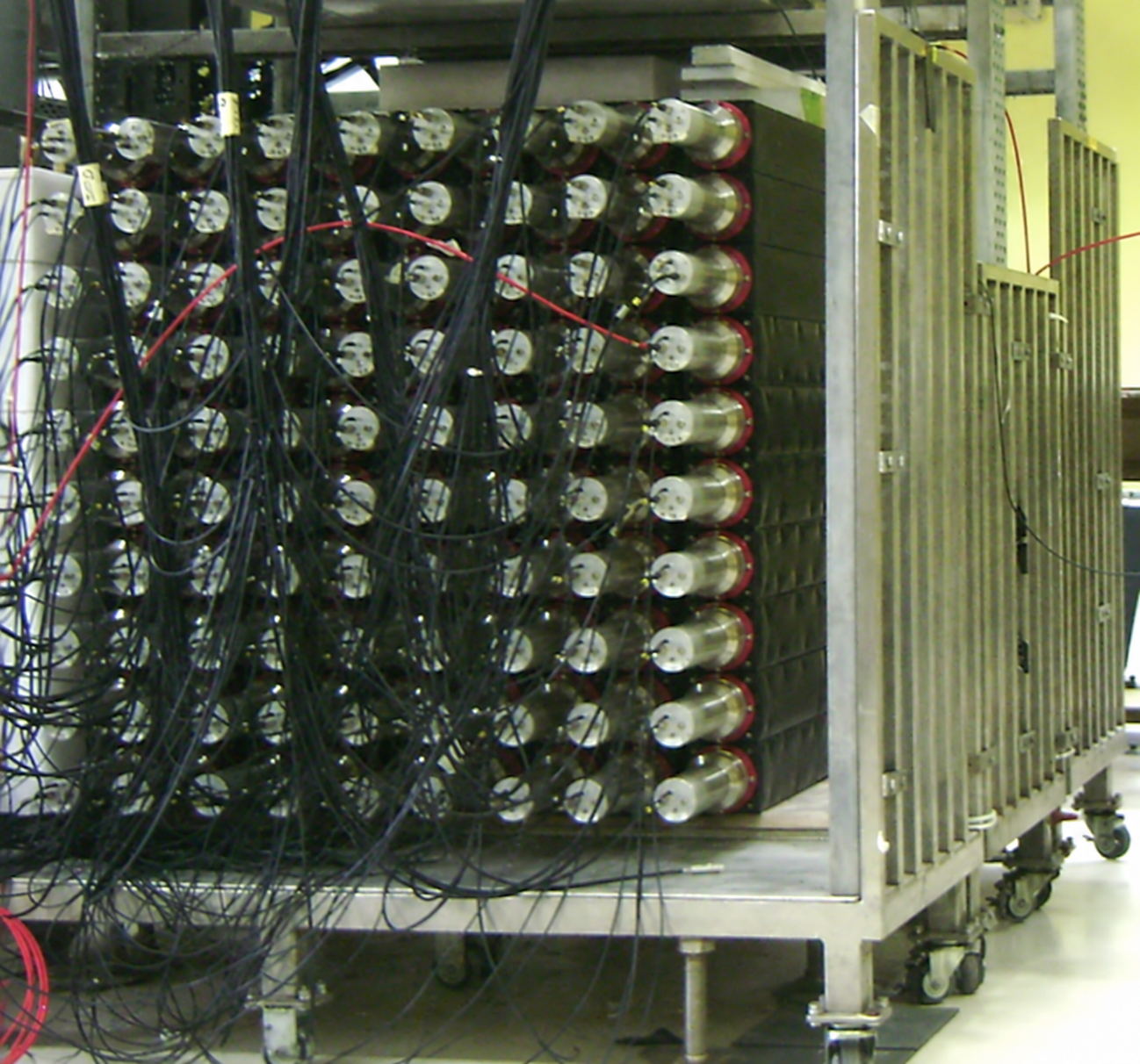}	
	\caption{The full-scale ISMRAN detector setup inside the Detector integrator laboratory.} 
	\label{fig_1}
\end{figure}

This approach is particularly relevant to the Indian Scintillator Matrix for Reactor Anti-Neutrinos (ISMRAN) experiment, which is designed to measure reactor ${\overline{\ensuremath{\nu}}}_{e}$ through IBD at very short baselines of 13-18~m from the Dhruva~\cite{dhruva} research reactor core using a one ton segmented plastic scintillator detector array~\cite{ismran_hql2023}, as shown in Fig.~\ref{fig_1}. Operating at the surface exposes ISMRAN to substantial cosmogenic and reactor-induced correlated backgrounds. Previous ISMRAN studies have shown that cosmic muon associated events can be substantially reduced using time coincidence technique~\cite{ismran_muon_roni}. Fast neutron induced correlated events, however, can survive the conventional IBD selection because their reconstructed energies, hit multiplicities, and prompt-delayed timing can overlap with those of genuine IBD interactions.

A fast neutron can undergo elastic scattering from a proton to produce a prompt recoil signal before being captured, thereby reproducing the characteristic prompt-delayed coincidence of an IBD interaction~\cite{DD_roni}. Although conventional selections based on energy, multiplicity, and timing provide significant background mitigation, they do not fully exploit the multidimensional information available in a segmented detector array. Consequently, achieving high background discrimination while maintaining high signal efficiency remains a challenging task.

Machine learning (ML) techniques have become increasingly important in particle and neutrino physics for event classification, reconstruction, and background suppression~\cite{juno_ml}. Rather than imposing independent thresholds on individual observables, a supervised classifier can learn the non-linear correlations among energy deposition and topology variables that characterize different interaction mechanisms. For a segmented scintillator detector, this approach is physically motivated by the distinct spatial and energy sharing patterns produced by electromagnetic positron interactions and proton recoils. The detailed segmentation of ISMRAN provides a particularly suitable detector response for topology-based classification~\cite{ismran_jinst}.

In this work, we develop a multilayer perceptron (MLP)-based event classification framework for prompt event identification in ISMRAN using reconstructed topological observables obtained from detailed \textsc{Geant4} simulations. The classifier performance is evaluated on an independent balanced MC reference sample, while the standard ISMRAN event selection is retained as a separate detector-level benchmark. Beyond event classification, the selected sample is propagated through a detector-level spectral likelihood to determine whether the underlying IBD normalization and reconstructed prompt energy spectrum can be recovered after explicitly accounting for the residual fast neutron component and the energy dependent classifier acceptance. The results demonstrate that topology-based classification can substantially suppress correlated fast neutron backgrounds while allowing the resulting energy-dependent signal acceptance to be propagated explicitly in the subsequent spectral likelihood for reactor ${\overline{\ensuremath{\nu}}}_{e}$ measurements. As a physics application, the MLP classifier score is further incorporated together with the reconstructed prompt energy in a joint likelihood for a two baseline sterile-neutrino sensitivity study at 13~m and 18~m, where the topology information provides an additional handle for constraining fast neutron backgrounds and improves the expected sensitivity relative to an energy-only analysis. This work therefore establishes a direct connection between topology-based background suppression and the sterile-neutrino physics potential of the ISMRAN experiment.

\section{IBD signature and benchmark event selection in ISMRAN}
\label{sec:IBD}

Reactor ${\overline{\ensuremath{\nu}}}_{e}$ are detected in ISMRAN through the IBD interaction,

\begin{equation}
\bar{\nu}_e + p \rightarrow e^{+} + n ,
\label{eq:ibd}
\end{equation}

which has a kinematic threshold of 1.806 MeV. The IBD signature consists of a prompt positron signal followed by a delayed neutron capture signal. The positron deposits its kinetic energy in the scintillator and subsequently annihilates, producing additional energy deposition from the resulting two 0.511 MeV annihilation photons. The neutron, in contrast, undergoes elastic scattering and thermalization before being captured after a characteristic delay. The combination of the prompt electromagnetic energy deposition and delayed neutron capture provides the characteristic time correlated signature used for IBD identification.

The ISMRAN detector consists of 90 plastic scintillator bars arranged in a $9\times10$ segmented array. Each scintillator bar is wrapped with $\mathrm{Gd_{2}O_{3}}$-coated aluminized Mylar foils to enhance neutron capture efficiency. The segmented geometry provides detailed event topology information through the spatial distribution of the deposited energy and the corresponding bar-hit multiplicity. In the detector geometry, $\sim$ 75\% of neutrons are captured on gadolinium (Gd), while the remaining 25\% are captured on hydrogen~\cite{ismran_dhruv}. Neutron capture on Gd produces a cascade of $\gamma$-rays with a total energy of $\sim$ 8 MeV and a mean capture time of $\sim65~\mu$s, providing a distinctive delayed signature for IBD selection.

Throughout this work, the performance of the proposed ML algorithm is compared with the standard ISMRAN cut-based event selection. Candidate IBD events are required to satisfy the following criteria:

\begin{itemize}

\item \textbf{Fiducial volume selection:}
Events with energy deposition in the outermost detector layer are rejected.

\item \textbf{Prompt energy:} Prompt candidates are required to satisfy $\mathrm{0.8 < E^{prompt}_{sum} (MeV) < 10.0}$, consistent with the expected visible energy range of reactor IBD positron events.

\item \textbf{Prompt multiplicity:} The prompt signal is required to have $\mathrm{2 \leq N_{bars}^{prompt} \leq 4}$, reflecting the compact energy deposition topology of positron-induced electromagnetic interactions in the segmented detector array.

\item \textbf{Delayed energy:} Delayed candidates are selected within $\mathrm{2.6 < E_{delayed}^{sum} (MeV) < 9.0}$, corresponding to the energy region populated by neutron-capture $\gamma$-rays. The lower threshold suppresses the 2.2 MeV $\gamma$-ray from neutron capture on hydrogen as well as reactor and environmental $\gamma$-ray backgrounds concentrated below $\sim$ 3 MeV, yielding a delayed sample dominated by Gd captures. 


\item \textbf{Delayed multiplicity:} The delayed event is required to satisfy $\mathrm{5 \leq N_{bars}^{delayed} \leq 10}$, consistent with the spatially extended topology of the $\gamma$-ray cascade following neutron capture on Gd.

\item \textbf{Time coincidence:} Prompt-delayed pairs are required to satisfy $\mathrm{8 < \Delta T_{PD} (\mu s) < 250}$, which is compatible with the neutron capture time distribution in ISMRAN and suppresses accidental coincidences.

\item \textbf{Vertex separation:}
The distance between the reconstructed prompt and delayed vertices is required to satisfy $\mathrm{\Delta R_{PD} < 0.5}$ m.

\item \textbf{Cosmic muon veto:}
Events occurring within $240~\mu$s of a tagged cosmic muon candidate are rejected~\cite{ismran_muon_roni}.

\item \textbf{Topology requirements:}
Additional topology-based selections are applied to suppress accidental and correlated backgrounds. In particular, both the prompt and delayed candidates are required to satisfy a pattern cut, in which the fired scintillator bars form contiguous clusters consistent with the expected IBD topology~\cite{ismran_prototype}.

\end{itemize}

\begin{figure}
	\centering 
	\includegraphics[width=0.75\textwidth]{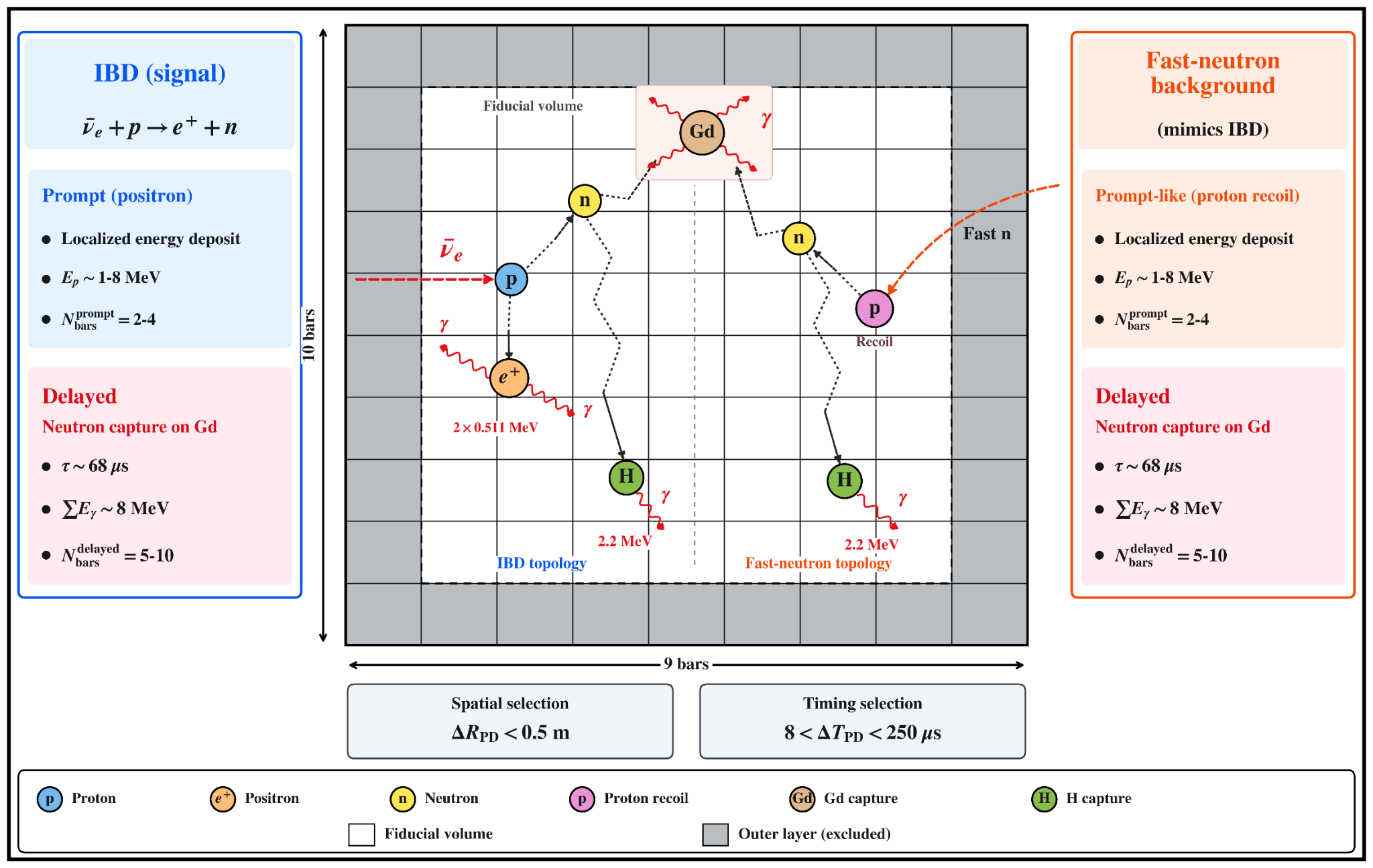}	
	\caption{Schematic representation of an IBD interaction and a fast neutron-induced background event in the ISMRAN detector. The segmented geometry provides topological information that can be exploited to distinguish positron-induced electromagnetic interactions from proton recoil events.} 
	\label{fig_2}
\end{figure}

These requirements define the conventional detector-level benchmark IBD selection. The reduced prompt level preselection used to construct the ML reference sample is defined separately in Sec.\ref{sec:sim1}. Figure~\ref{fig_2} illustrates the characteristic topologies of reactor IBD and fast neutron-induced background events in the full-scale ISMRAN detector array.


The benchmark preselection provides effective suppression of accidental and correlated backgrounds, however, the sequential selection criteria do not fully exploit the correlations among reconstructed detector observables. In particular, electromagnetic energy deposition pattern and fast neutron-induced proton recoils exhibit overlapping distributions in the multidimensional feature space, limiting the achievable background rejection while maintaining high IBD efficiency. These limitations motivate the use of ML techniques capable of combining multiple topological observables simultaneously, developed later in this work. 

\section{Expected backgrounds in reactor ${\overline{\ensuremath{\nu}}}_{e}$ measurements}
\label{sec:bkg}

Reactor ${\overline{\ensuremath{\nu}}}_{e}$ measurements based on the IBD interaction are affected by both accidental and correlated backgrounds. In above-ground very short baseline experiments such as ISMRAN, the two classes require different suppression strategies. Accidental backgrounds arise from statistically independent detector signals, whereas correlated backgrounds originate from a common physical interaction and can reproduce the characteristic prompt-delayed IBD signature.

\textbf{Accidental backgrounds} result from random coincidences between uncorrelated prompt-like and delayed-like signals. Their principal sources include environmental $\gamma$ radiation, and natural radioactivity. Because the two signals are uncorrelated in time, the accidental contribution can be estimated using off-time coincidence windows, while reactor-OFF data provide an independent characterization of non-reactor backgrounds~\cite{ismran_prototype}. These backgrounds can therefore be constrained statistically and are not the primary focus of the present study.

\textbf{Correlated fast neutron backgrounds} are more challenging because the prompt and delayed signals arise from the same neutron interaction chain. Fast neutrons produced by cosmic-ray interactions~\cite{ismran_muon_roni} or reactor-related processes can enter the detector and scatter elastically from protons, producing a prompt proton recoil signal. The neutron may subsequently thermalize and be captured on Gd or H nuclei, generating a delayed capture signal. The reconstructed prompt energy, bar multiplicity, spatial topology, and prompt-delayed timing can therefore overlap substantially with those of true IBD interactions.

The benchmark preselection described in Sec.~\ref{sec:IBD} suppresses a large fraction of these events, but residual fast neutron backgrounds remain because proton recoil and positron-induced electromagnetic energy deposition patterns are not completely separable using sequential one-dimensional requirements. The remaining discrimination is therefore expected to reside primarily in correlations among the reconstructed prompt topology observables. This motivates the multivariate classifier introduced in Sec.~\ref{sec:ml}, which is designed to exploit these correlations for fast neutron rejection.

\section{Monte Carlo simulation and dataset preparation}
\label{sec:sim}

A detailed \textsc{Geant4}-based Monte Carlo (MC) framework was used to generate the IBD signal and fast neutron background samples employed in this study~\cite{Geant4}. The simulation incorporates the full ISMRAN detector geometry, particle transport, energy deposition, neutron thermalization and capture, and the detector response model used in the standard ISMRAN reconstruction chain~\cite{ismran_dhruv,ismran_roni_nima}.

\textbf{IBD dataset:}
A total of $10^{6}$ reactor ${\overline{\ensuremath{\nu}}}_{e}$ interactions were simulated uniformly and isotropically throughout the detector volume. The incident ${\overline{\ensuremath{\nu}}}_{e}$ spectrum was constructed from the reactor isotope dependent flux model ~\cite{Mueller,Huber} and fission fractions adopted for the Dhruva research reactor~\cite{dhruva}, together with the corresponding IBD cross section~\cite{Vogel}. The positron and neutron kinematics were generated according to the IBD interaction, followed by full particle transport and detector response simulation. Electromagnetic interactions were treated using the \textsc{Geant4} electromagnetic physics models, while hadronic transport was described using the QGSP\_BIC\_HP physics list. The DICEBOX package~\cite{dicebox} was used to model the statistical $\gamma$-ray cascades following neutron capture on Gd~\cite{ismran_roni_nima}.

\textbf{Fast neutron dataset:}
A total of $10^{6}$ fast neutrons were generated above the detector shielding with kinetic energies uniformly distributed between $2$ and $12~\mathrm{MeV}$ and propagated downward-isotropic through the complete detector geometry using the same transport and reconstruction framework as the IBD sample. The uniform spectrum is not intended to represent a measured reactor neutron energy distribution, rather, it provides broad sampling of the neutron energies capable of producing prompt-like proton recoil topologies in ISMRAN. The dependence of the classifier performance on this assumed training spectrum is examined independently using an AmBe neutron spectrum in Sec.~\ref{ambe}.

\begin{figure*}[t]
\centering
\includegraphics[width=1.0\textwidth]{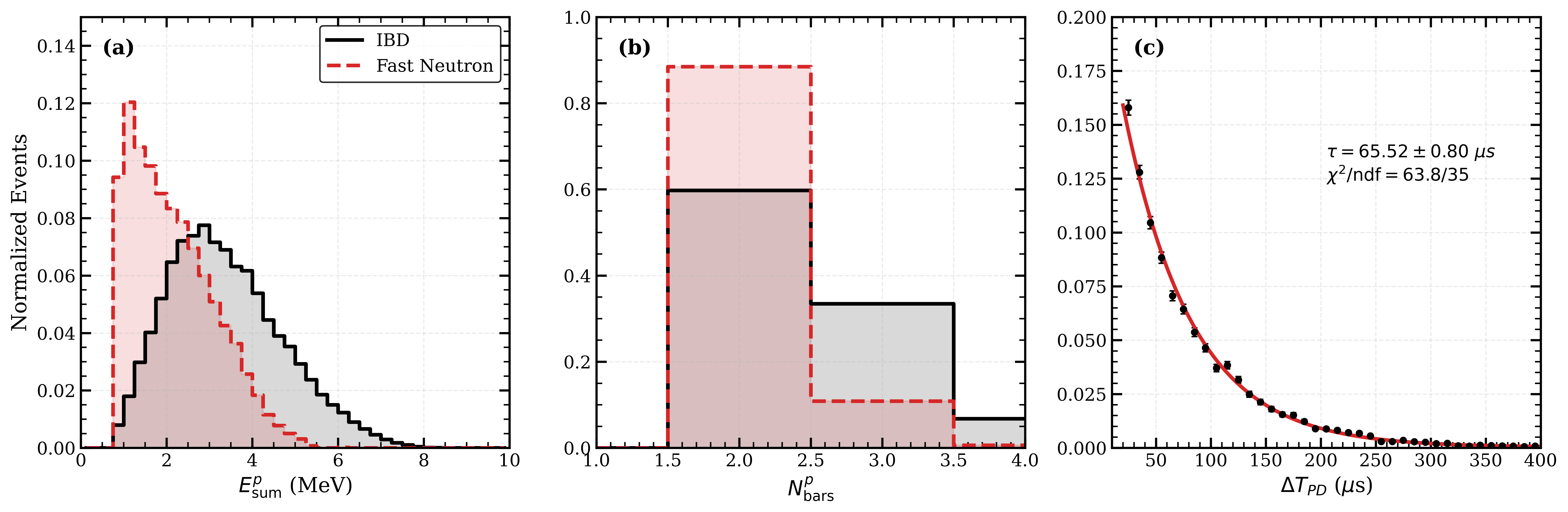}
\caption{Primary observables used for IBD event reconstruction in ISMRAN. (a) Reconstructed prompt energy distributions for simulated IBD and fast neutron events. (b) Prompt scintillator-bar multiplicity for the two event classes. (c) Prompt-delayed time difference for simulated IBD events. The solid red curve shows the exponential plus constant fit used to determine the neutron-capture time.}
\label{fig_3}
\end{figure*}

\subsection{Mixed signal-background dataset}
\label{sec:sim1}

Detector response effects were incorporated by applying the measured energy and position resolutions to the simulated energy deposits before event reconstruction~\cite{ismran_roni_nima}. For the classifier study, a common prompt-level reference population is defined by requiring $\mathrm{2 \leq N_{bars}^{prompt} \leq 4}$, and $\mathrm{0.8 < E^{prompt}_{sum} (MeV) < 10.0}$. No classifier requirement is applied at this stage. For the ML study, equal numbers of reconstructed IBD and fast neutron events passing this common pre-selection are randomly sampled to construct an exactly balanced $1:1$ reference dataset. This dataset was subsequently divided into statistically independent training, validation, and testing samples. The balanced class composition is adopted to prevent the optimization from being driven by an arbitrary difference in the available MC sample sizes and to provide a controlled reference sample for classifier comparison. It does not represent the physical IBD to fast neutron event rate ratio expected in ISMRAN. Consequently, quantities that depend on class composition, such as purity, $S/B$, and the $F_1$ score, are interpreted as benchmark classification metrics for this reference sample unless physical signal and background weights are explicitly introduced.

Figure~\ref{fig_3} illustrates the reconstructed quantities entering the benchmark IBD selection. Figures~\ref{fig_3}(a) and \ref{fig_3}(b) show the prompt summed energy, $\mathrm{E^{p}_{sum}}$, and prompt bar multiplicity, $\mathrm{N^{p}_{bars}}$, for simulated IBD and fast neutron events. The strong overlap of the prompt energy distributions illustrates why an energy selection alone provides limited discrimination, while the different multiplicity distributions already reflect differences in the spatial development of positron and proton recoil energy deposition. Figure~\ref{fig_3}(c) shows the prompt-delayed time distribution, $\mathrm{\Delta T_{PD}}$, for reconstructed IBD candidates. An exponential plus constant description reproduces the expected neutron-capture time profile and yields a characteristic capture time consistent with previous ISMRAN measurements~\cite{ismran_prototype}.

The benchmark cut-based selection provides the reference detector-level performance against which the subsequent multivariate selection is assessed. Its absolute detector-level efficiency relative to generated IBD is defined as,
\begin{equation}
\epsilon_{\rm IBD}^{\rm bench}= \frac{N_{\rm IBD}^{\rm bench}}{N_{\rm IBD}^{\rm generated}},
\label{eq:ibd_efficiency_benchmark}
\end{equation}

where $\mathrm{N_{IBD}^{bench}}$ is the number of simulated IBD interactions surviving detector reconstruction and the complete benchmark cut-based selection, and $\mathrm{N_{IBD}^{generated}}$ is the number of generated IBD interactions in the corresponding reference volume. This quantity therefore includes reconstruction acceptance and all benchmark event selection requirements.

For a selected signal-background sample, the IBD purity is defined as
\begin{equation}
f_{IBD} = \frac{\mathrm{N_{IBD}^{selected}}}{\mathrm{N_{IBD}^{selected}+N_{FN}^{selected}}},
\label{eq:ibd_purity}
\end{equation}

where $\mathrm{N_{FN}^{selected}}$ denotes the number of fast neutron events surviving the same selection. The purity corresponds to the precision of the IBD class in binary classification. Unlike the IBD efficiency, however, it depends explicitly on the relative signal and background populations. Purity values obtained from the balanced MC reference sample should therefore not be interpreted as the expected reactor-ON purity of the experimental ISMRAN dataset.

For completeness, the benchmark sample is also characterized using the counting-based figure of merit
\begin{equation}
\mathcal{F}
=
\frac{S}{\sqrt{S+B}} .
\label{eq:benchmark_fom}
\end{equation}

Because its numerical value depends on both the size and the signal to background composition of the simulated sample, $\mathcal{F}$ is used only as a comparative benchmark and is not interpreted as the expected experimental significance or sensitivity of ISMRAN.

\begin{table}[ht]
\centering
\caption{Performance of the benchmark cut-based IBD selection in the simulated IBD and fast neutron reference sample. The quantities depending on sample composition are benchmark MC metrics and should not be interpreted as reactor-ON experimental values.}

\label{tab:cutbased_performance}
\begin{tabular}{lc}
\hline
Quantity & Value \\
\hline
IBD absolute efficiency, $\epsilon_{\rm IBD}^{\rm bench}$ & 37.7\% \\
Signal purity, $f_{\rm IBD}$ & 29.9\% \\
Signal to background ratio, $S/B$ & 0.426 \\
Benchmark figure of merit, $\mathcal{F}$ & 36.1 \\
\hline
\end{tabular}
\end{table}

These definitions are used consistently throughout the analysis. The signal efficiency and purity are used throughout the analysis as complementary measures of event selection performance. The signal efficiency quantifies the acceptance of genuine IBD interactions for the specified reference population, whereas the purity characterizes the composition of the selected candidate sample. Both quantities are therefore required to assess the performance of an event selection strategy for reactor ${\overline{\ensuremath{\nu}}}_{e}$ measurements.

\section{Machine learning framework}
\label{sec:ml}

The benchmark selection combines reconstructed energy, multiplicity, timing, spatial-coincidence, and topology requirements to identify IBD candidates. Although these observables provide substantial background suppression, sequential selections do not fully exploit correlations among the prompt event quantities retained by the segmented ISMRAN detector. In particular, positron-induced electromagnetic energy deposition and fast neutron-induced proton recoils can overlap in individual observables while remaining distinguishable through their combined energy sharing and spatial characteristics.

A supervised multilayer perceptron (MLP) is therefore used to combine three reconstructed prompt topology observables into a single event-level discriminant. IBD events are assigned the target label $y=1$, whereas fast neutron events are assigned $y=0$. The classifier is trained on the balanced MC reference dataset described in Sec.~\ref{sec:sim}. The balanced class composition controls the relative contribution of the two classes during optimization but is not intended to reproduce their physical event rate ratio in ISMRAN.

The MLP output is denoted by $\mathrm{P_{IBD}}$ and lies between zero and one. Larger values indicate prompt topologies that are more signal-like within the simulated reference samples. The score is used as a classifier discriminant rather than interpreted as a calibrated physical probability. Its operating threshold is determined using the validation sample and is subsequently fixed before evaluating the independent test sample and the likelihood-based signal extraction described in Sec.~\ref{sec:likelihood}.

\subsection{Input observables}
\label{sec:features}

The MLP uses three reconstructed observables that probe complementary aspects of the prompt energy deposition topology: the concentration of energy in the leading scintillator bars, the non-uniformity of energy sharing among the fired bars, and the magnitude of the leading energy deposit relative to the prompt multiplicity. All observables are constructed after applying the detector response model and standard ISMRAN reconstruction, so that the classifier acts on detector-level quantities rather than truth-level information~\cite{ismran_roni_nima}.

The physical motivation arises from the different energy-loss and transport mechanisms of positrons and recoil protons in plastic scintillator. An IBD positron loses energy through ionization and electromagnetic interactions and subsequently annihilates, producing two 511~keV $\gamma$ rays whose secondary interactions can redistribute energy among neighbouring scintillator bars. A recoil proton produced by fast neutron elastic scattering has a shorter range and a larger specific energy loss, leading to a different spatial and energy sharing pattern at the bar level~\cite{knoll}. The segmentation of ISMRAN preserves these differences at the reconstructed event level and therefore provides observables sensitive to the spatial concentration and energy-sharing pattern of the prompt cluster.

The three observables are constructed from the reconstructed prompt energies of the fired scintillator bars. Their definitions and physical interpretations are given below. Before training, each observable is standardized to zero mean and unit variance using only the corresponding statistics of the training sample. The same transformation is then applied unchanged to the validation and testing samples.

\begin{figure*}[t]
	\centering 
	\includegraphics[width=1.0\textwidth]{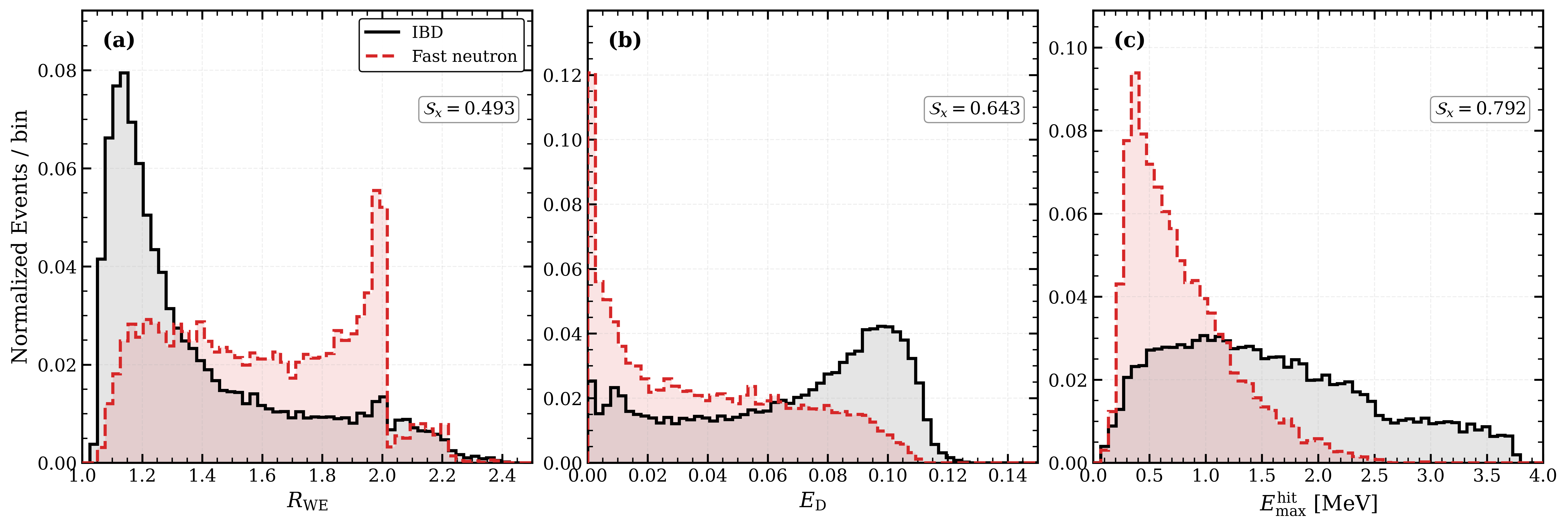}	
	\caption{Distributions of the three reconstructed prompt-topology observables used as inputs to the MLP classifier for simulated IBD and fast neutron events: (a) weighted energy ratio $\mathrm{R_{WE}}$,(b) energy dispersion $\mathrm{E_{D}}$, and (c) maximum-bar energy observable $\mathrm{E^{hit}_{max}}$. The distributions are obtained after the event reconstruction and preselection requirements applied to the ML reference sample.} 
	\label{fig_4}
\end{figure*}

\textbf{(A) Weighted energy ratio ($\mathrm{R_{WE}}$)}:
The weighted energy ratio ($\mathrm{R_{WE}}$) quantifies the contribution of the two most energetic scintillator bars relative to the overall energy-weighted topology of the prompt cluster. It is defined as

\begin{equation}
R_{\rm WE}
=
\frac{E_{\rm max}+E_{\rm 1st}}
{W_{\rm sum}},
\label{eq:RWE}
\end{equation}

where $E_{\rm max}$ and $E_{\rm 1st}$ denote the energies deposited in the highest and second-highest energy scintillator bars, respectively, and

\begin{equation}
W_{\rm sum}
=
\sum_{i=1}^{N_{\rm bars}}
\left(
\frac{E_i}{E_{\rm sum}}
\right)E_i
=
\frac{\sum_{i=1}^{N_{\rm bars}}E_i^2}
{E_{\rm sum}},
\label{eq:Wsum}
\end{equation}

The quantity $\mathrm{W_{sum}}$ is an energy-weighted measure of the prompt energy distribution, with larger deposited energies receiving greater weight. The ratio $\mathrm{R_{WE}}$ therefore characterizes how strongly the two leading bar deposits contribute relative to the overall energy weighted topology. As shown in Fig.~\ref{fig_4}(a), the IBD and fast neutron samples populate this observable differently, reflecting their different prompt energy sharing patterns. The observable therefore provides one component of the multivariate signal background discrimination.

\textbf{(B) Energy dispersion ($\mathrm{E_{D}}$)}:
The energy dispersion ($\mathrm{E_{D}}$) observable quantifies the non-uniformity of energy sharing among the fired scintillator bars and is defined as

\begin{equation}
E_{\rm D}
=
\sum_{i=1}^{N_{\rm bars}}
\frac{E_i}{E_{\rm sum}}
\left(
\frac{E_i}{E_{\rm sum}}
-
\frac{1}{N_{\rm bars}}
\right)^2,
\label{eq:ED}
\end{equation}

where $\mathrm{E_{i}}$ is the energy deposited in the $i$th scintillator bar and $\mathrm{N_{bars}}$ is the prompt bar multiplicity. The observable compares the measured energy fractions with an equal-sharing configuration, approaching zero for a uniform distribution and increasing as the energy becomes more unevenly distributed.

Unlike $\mathrm{R_{WE}}$, which is particularly sensitive to the leading energy deposits, $\mathrm{E_{D}}$ incorporates the energy fractions of all fired bars. It approaches zero for approximately uniform energy sharing and increases as the distribution becomes more asymmetric. The different $\mathrm{E_{D}}$ distributions observed for the simulated IBD and fast neutron samples in Fig.~\ref{fig_4}(b) therefore provide complementary information on the global prompt energy-sharing topology.

\textbf{(C) Maximum-bar energy observable ($\mathrm{E^{hit}_{max}}$)}:
The third observable is defined as

\begin{equation}
E^{hit}_{\rm max}
=
\frac{E_{\rm max}}
{N_{\rm bars}}.
\label{eq:EM}
\end{equation}

The observable combines the magnitude of the leading localized energy deposit with the reconstructed prompt multiplicity. Its distributions for IBD and fast neutron events, shown in Fig.~\ref{fig_4}(c), exhibit the largest individual separation among the three input variables considered here. It therefore provides complementary information to the energy-sharing observables $\mathrm{R_{WE}}$ and $\mathrm{E_{D}}$.

The three observables consequently probe complementary aspects of the prompt topology: the relative contribution of the leading energy deposits through $\mathrm{R_{WE}}$, the global energy-sharing pattern through $\mathrm{E_{D}}$, and the localized energy concentration through $\mathrm{E_{max}^{hit}}$. Their usefulness is enabled by the segmentation of ISMRAN, which retains spatial information on the energy deposition at the scintillator bar level. To quantify the one-dimensional discrimination provided by each observable, we define the dimensionless mean-separation metric

\begin{equation}
\mathcal{S}_{x} =
\frac{
\left|\mu_{\rm IBD}-\mu_{\rm FN}\right|
}{
\sqrt{
\sigma_{\rm IBD}^{2}+\sigma_{\rm FN}^{2}
}
},
\label{eq:separation_metric}
\end{equation}

where $\mathrm{\mu_{IBD}}$ and $\mathrm{\mu_{FN}}$ are the sample means of observable $x$, and $\mathrm{\sigma_{IBD}}$ and $\mathrm{\sigma_{FN}}$ are the corresponding RMS widths. The metric compares the displacement of the two distribution centroids with their combined event-by-event spread. It is used only as a descriptive one-dimensional measure and is not interpreted as a classification efficiency or statistical significance.

For the distributions shown in Fig.~\ref{fig_4}, the resulting values are $\mathcal{S}_{R_{\rm WE}}=0.49$, $\mathcal{S}_{E_D}=0.64$, and $\mathcal{S}_{E_{\max}^{\rm hit}}=0.79$. The $\mathrm{E_{max}^{hit}}$ observable therefore provides the largest individual separation, while $\mathrm{R_{WE}}$ and $\mathrm{E_{D}}$ retain complementary information on the leading energy deposits and the overall inter-bar energy sharing. Their simultaneous use in the MLP allows correlations among these complementary topology observables to be exploited beyond their individual one-dimensional separation.


\subsection{MLP classifier architecture and loss function}
\label{sec:mlp}

The MLP is a fully connected feed-forward network with three input nodes corresponding to $\mathrm{R_{WE}}$, $\mathrm{E_{D}}$, and $\mathrm{E_{max}^{hit}}$. Four hidden layers containing 128, 64, 32, and 16 neurons are used, each with a rectified linear unit (ReLU) activation. Batch normalization is applied after the first two hidden layers. Dropout with rates of 0.25, 0.25, 0.20, and 0.20 is applied after the successive hidden layers, while an $\mathrm{L_2}$ penalty of $\mathrm{10^{-4}}$ is applied to the kernel weights of the first three dense layers. These regularization procedures reduce sensitivity to statistical fluctuations in the training sample and limit overfitting.

The output layer contains a single neuron with sigmoid activation,

\begin{equation}
P_{IBD}
=
\frac{1}{1+\exp(-z)},
\label{eq:sigmoid}
\end{equation}

where $z$ denotes the pre-activation output of the final layer. $\mathrm{P_{IBD}}$ is used as a continuous classifier score: values closer to unity correspond to more IBD-like reconstructed topologies, whereas values closer to zero correspond to more fast neutron like topologies. No independent probability calibration procedure is applied, and the output is therefore interpreted as a discriminating score rather than a physical event probability.

The network parameters are optimized by minimizing the binary cross-entropy loss,

\begin{equation}
\mathcal{L}
=
-\frac{1}{N}
\sum_{i=1}^{N}
\left[
y_i\ln P_i
+
(1-y_i)\ln(1-P_i)
\right],
\label{eq:loss_func}
\end{equation}

where $\mathrm{P_{i}}$ is the MLP score for event $i$ and $\mathrm{y_{i}}=1$ ($0$) denotes an IBD (fast neutron) event. Minimization of this loss determines a non-linear decision boundary in the three-dimensional observable space.

\begin{figure}
	\centering 
	\includegraphics[width=0.9\textwidth]{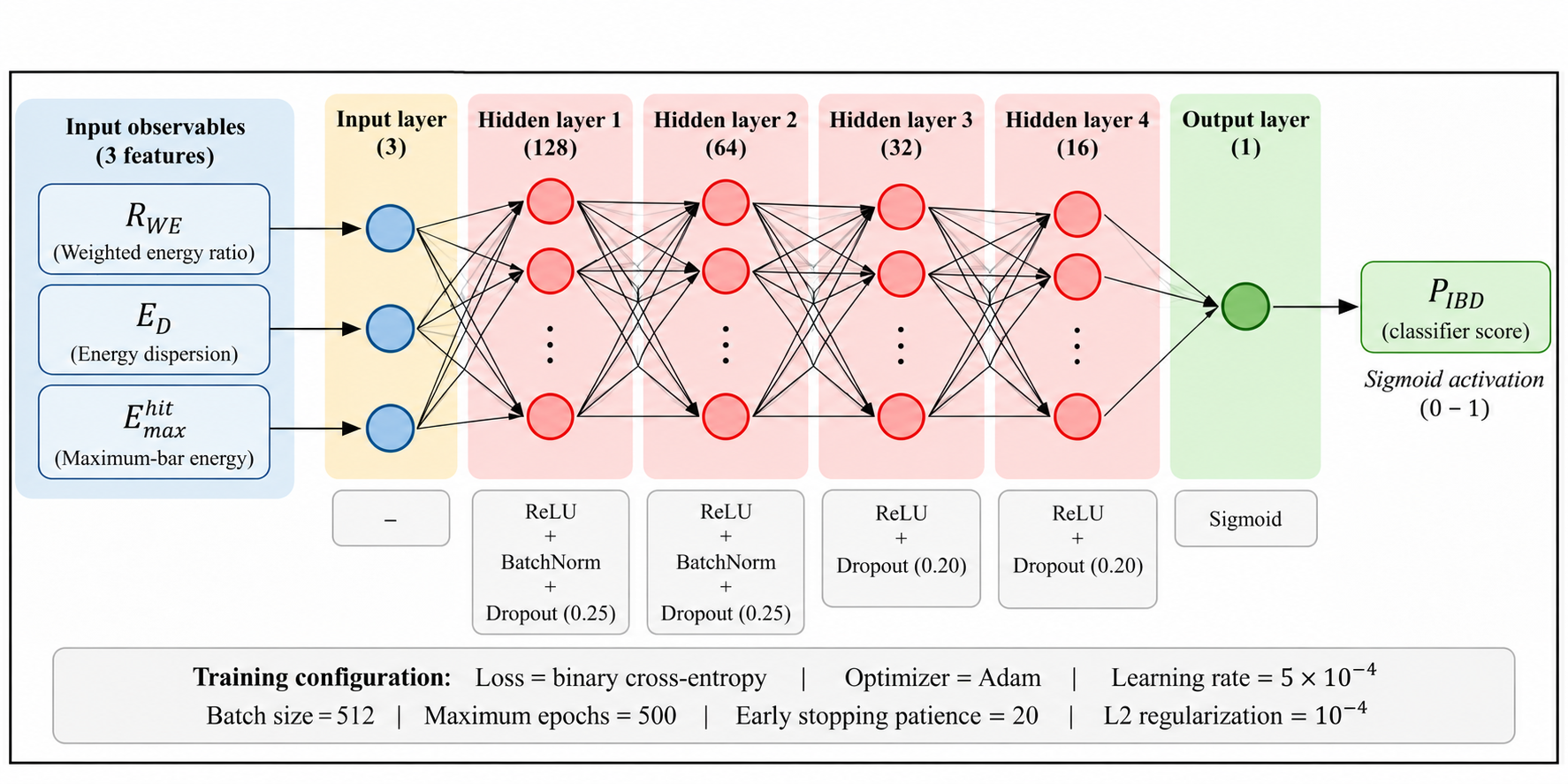}	
	\caption{Architecture of the final multilayer perceptron classifier. Three reconstructed topology observables are propagated through four fully connected hidden layers containing 128, 64, 32, and 16 neurons with ReLU activations. Batch normalization and dropout are used as indicated. The sigmoid output provides the classifier score $\mathrm{P_{IBD}}$.} 
	\label{fig_5}
\end{figure}

\begin{table}[t]
\centering
\caption{Architecture and training hyperparameters of the final MLP classifier.}
\label{tab:mlp}
\begin{tabular}{lc}
\hline
Parameter & Value \\
\hline
Input variables & 3 \\
Hidden layers & 4 \\
Neurons per hidden layer & 128, 64, 32, 16 \\
Hidden activation & ReLU \\
Output activation & Sigmoid \\
Batch normalization & First two hidden layers \\
Dropout rates & 0.25, 0.25, 0.20, 0.20 \\
$L_2$ regularization & $10^{-4}$ \\
Loss function & Binary cross-entropy \\
Optimizer & Adam \\
Initial learning rate & $5\times10^{-4}$ \\
Batch size & 512 \\
Maximum epochs & 500 \\
Early-stopping patience & 20 epochs \\
Trainable parameters & 11,777 \\
Classifier output & $P_{\rm IBD}$ \\
\hline
\end{tabular}
\end{table}

\subsection{Training, validation and threshold procedure}
\label{sec:training}

The balanced reference sample is randomly divided into statistically independent training, validation, and testing subsets. In the final analysis, 11,070 events are used for training, 3,691 for validation, and 6,327 for testing, corresponding to $\sim$ 52.5\%, 17.5\%, and 30.0\% of the selected reference sample, respectively. Stratified sampling is used to preserve the equal signal-background composition in each subset. The training sample is used to optimize the network parameters, the validation sample is used for early stopping and operating threshold determination, and the testing sample is reserved for the final performance evaluation.

The Adam optimizer with a learning rate of $\mathrm{5\times10^{-4}}$ is used to minimize the binary cross-entropy loss. Training is performed with a batch size of 512 for at most 500 epochs. Early stopping is applied when the validation loss fails to improve by at least $\mathrm{10^{-5}}$ for 20 consecutive epochs, and the network parameters corresponding to the minimum validation loss are restored.

After training, the continuous classifier score is converted into an event selection by applying a threshold. The operating point is determined from the validation sample by maximizing the $\mathrm{F_1}$ score, which balances IBD efficiency and signal purity for the balanced reference sample. The statistical stability of the optimized threshold is evaluated using bootstrap resampling, as described in Sec.~\ref{sec:roc}. Once determined, the threshold is frozen and applied without further optimization to the independent testing sample.

\begin{figure*}[t]
\centering
\includegraphics[width=1.0\textwidth]{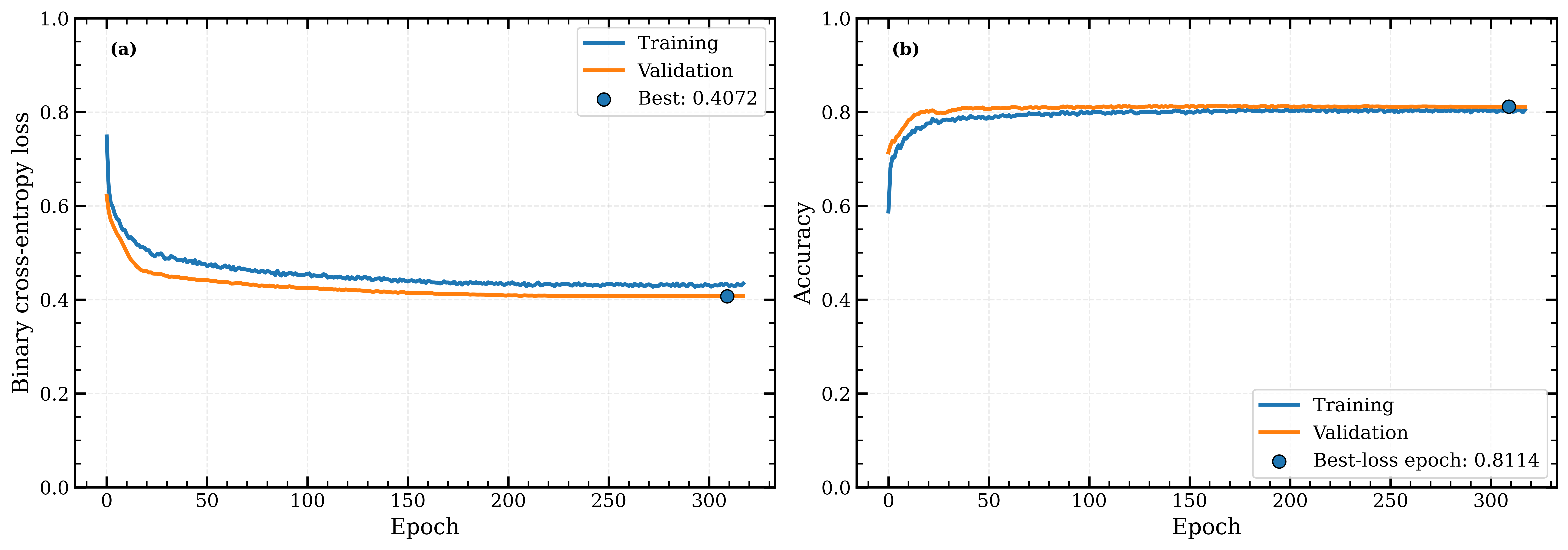}
\caption{(a) Training and validation binary cross-entropy loss and (b) classification accuracy as functions of epoch. The marker denotes the checkpoint selected by the minimum validation loss.}
\label{fig_6}
\end{figure*}

Figure~\ref{fig_6} shows the evolution of the training and validation losses during optimization. Both decrease during the initial optimization and approach similar stable values before early stopping. The absence of a sustained decrease in the training loss accompanied by an increasing validation loss provides no clear indication of substantial overfitting in this training configuration.

\subsection{Benchmark classifiers}
\label{sec:benchmark_classifiers}

For comparison, two additional classifiers are trained using the same three standardized input observables and the same statistically independent training and testing samples. The baseline fully connected neural network (FCNN) contains hidden layers of 64, 32, and 16 neurons with ReLU activations and dropout regularization. A gradient boosted decision tree (BDT) classifier is also trained with 300 estimators, a learning rate of 0.05, maximum tree depth of four, a minimum split size of five events, and a subsampling fraction of 0.9. These classifiers are used only as performance benchmarks. The MLP described above is used for the subsequent event selection and likelihood analysis.

The resulting MLP maps the three reconstructed topology observables onto a single event-level score. Its discrimination performance, threshold optimization, and comparison with the benchmark classifiers are presented in Sec.~\ref{sec:results}, while its propagation into the reactor ${\overline{\ensuremath{\nu}}}_{e}$ signal-extraction framework is discussed in Sec.~\ref{sec:likelihood}.

\section{Results and performance}
\label{sec:results}

The performance of the MLP-based event selection is evaluated using statistically independent validation and testing samples. The validation sample is used to determine the classifier operating threshold and to assess its statistical stability, whereas the testing sample is reserved for the final evaluation of the fixed classifier selection.

The analysis proceeds in four stages. First, the discrimination performance and the operating threshold are quantified using ROC curves and threshold dependent classification metrics. Second, the corresponding MLP-score distributions are examined to interpret the signal background separation. Third, the robustness of the classifier against a change in the fast neutron energy spectrum is tested. Finally, the reconstructed prompt energy spectrum is studied to assess possible spectral distortions introduced by the topology-based selection.

\subsection{Signal efficiency and background rejection}
\label{sec:roc}

\begin{figure*}[t]
\centering
\includegraphics[width=1.0\textwidth]{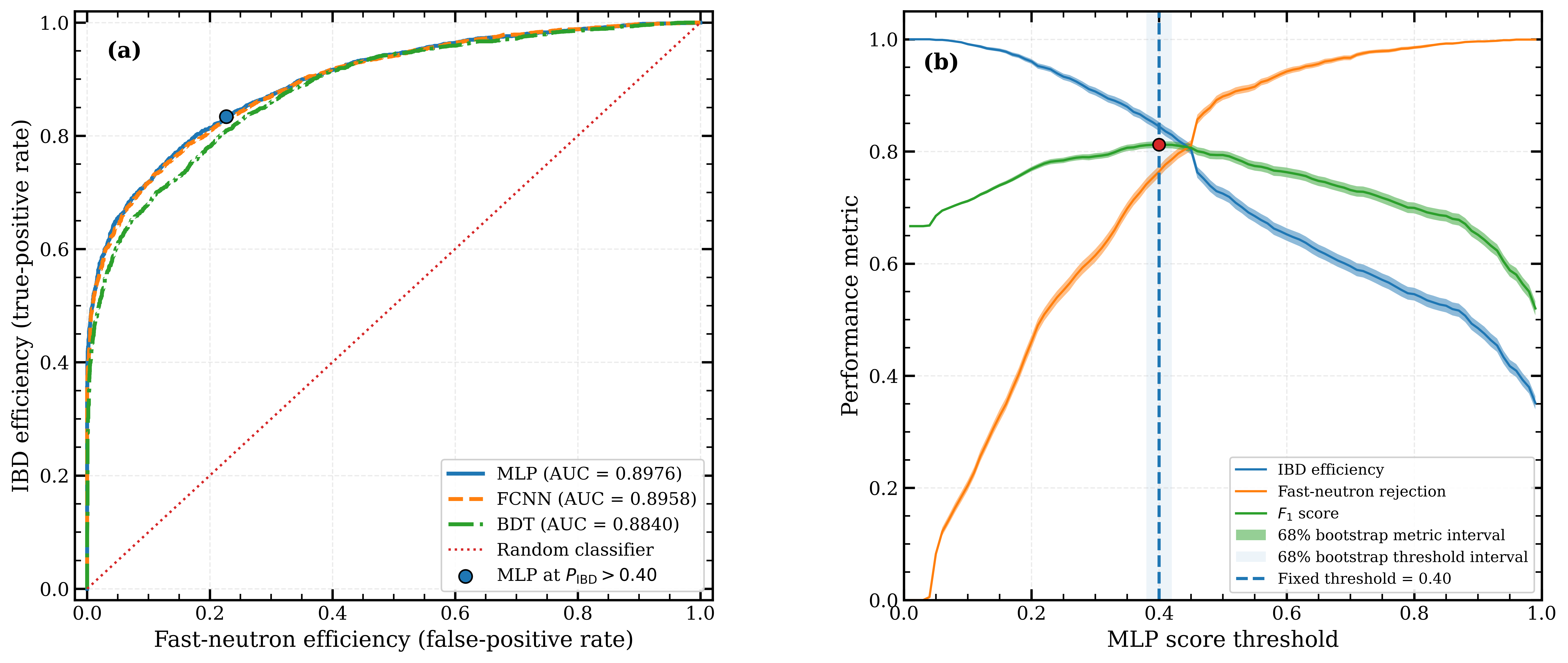}
\caption{Performance of the topology-based classifiers. (a) Receiver operating characteristic curves for the MLP, baseline FCNN, and BDT evaluated on the independent testing sample. The corresponding AUC values are 0.8976, 0.8958, and 0.8840, respectively. The marker indicates the final MLP operating point obtained by applying the threshold determined from the validation sample. (b) IBD signal efficiency, fast neutron rejection, and $\mathrm{F_{1}}$ score as functions of the MLP-score threshold evaluated on the validation sample. The shaded bands show the central 68\% bootstrap intervals of the threshold dependent metrics, while the vertical shaded region denotes the central 68\% bootstrap interval of the $\mathrm{F_{1}}$ optimal threshold.}
\label{fig:mlp_performance}
\end{figure*}

The discrimination performance is summarized in Fig.~\ref{fig:mlp_performance}. Figure~\ref{fig:mlp_performance}(a) shows receiver operating characteristic (ROC) curves for the MLP, baseline fully connected neural network (FCNN), and boosted decision tree(BDT), evaluated on the independent testing sample. The ROC curve gives the IBD signal efficiency (true positive rate) as a function of the fast neutron acceptance (false positive rate) as the classifier threshold is varied and therefore provides a threshold independent measure of signal background discrimination.

The MLP achieves an area under the ROC curve of $\mathrm{0.8976}$, compared with $\mathrm{0.8958}$ for the FCNN and $\mathrm{0.8840}$ for the BDT. Since all classifiers use the same reconstructed observables, $\mathrm{R_{WE}}$, $\mathrm{E_{D}}$, and $\mathrm{E_{max}^{hit}}$, provide substantial discrimination between IBD positrons and fast neutron-induced proton recoils. The MLP and FCNN network architectures therefore provide nearly identical nominal discrimination, while the BDT gives a slightly lower AUC. No statistical significance is assigned to the differences among these benchmark classifiers. 

The threshold dependence of the MLP performance is shown in Fig.~\ref{fig:mlp_performance}(b). As the MLP-score threshold is increased, the selection becomes progressively more restrictive: the IBD efficiency decreases, while the fast neutron rejection increases. The $\mathrm{F_{1}}$ score provides a convenient measure for selecting an operating point that balances signal acceptance and sample purity. It is defined as

\begin{equation}
F_1 =
\frac{2 f_{IBD}\epsilon_{IBD}^{ MLP}}
     {f_{IBD}+\epsilon_{IBD}^{MLP}},
\label{eq:f1_score}
\end{equation}

where $f_{IBD}$ is the IBD purity and $\mathrm{\epsilon_{IBD}^{MLP}}$ is the fraction of IBD events in the classifier reference sample that pass the MLP threshold ($\mathrm{P_{IBD}}>0.40$). Because $f_{IBD}$ depends on the assumed class composition, the $\mathrm{F_{1}}$ optimization defines the operating point for the balanced validation sample and should not be interpreted as a unique optimum for the physical reactor-ON signal to background ratio.

The operating threshold is determined exclusively from the validation sample. The nominal $\mathrm{F_1}$ maximum occurs at $\mathrm{P_{IBD}}=0.40$. To quantify the statistical stability of this operating point, the validation signal and background score distributions are independently resampled with replacement in 1000 bootstrap realizations. For each realization, the threshold that maximizes the $\mathrm{F_{1}}$ score is recorded. The resulting distribution has a median of 0.40 and a central 68\% interval of $[0.38,\,0.42]$. The fixed event-selection requirement is therefore

\begin{equation}
P_{\rm IBD}>0.40 .
\label{eq:mlp_threshold}
\end{equation}

Applying this frozen threshold to the independent testing sample gives an IBD MLP classification efficiency of $\mathrm{(83.4\pm 0.7)\%}$ and a fast neutron rejection of $\mathrm{(77.3\pm 0.7)\%}$, where the quoted uncertainties are binomial statistical uncertainties associated with the finite testing sample. The corresponding purity in the balanced testing sample is $\mathrm{78.61\%}$, with an $\mathrm{F_1}$ score of $\mathrm{0.809}$. The test sample operating point is indicated by the marker in Fig.~\ref{fig:mlp_performance}(a). Importantly, the testing sample is not used to optimize or modify the classifier threshold, providing an unbiased evaluation of the final selection performance. The resulting operating point therefore retains most of the IBD candidates entering the classifier while rejecting $\sim$ three quarters of the fast neutron events in the balanced testing sample.

\subsection{MLP score distribution}
\label{sec:mlp_score}

\begin{figure*}[t]
\centering
\includegraphics[width=0.85\textwidth]{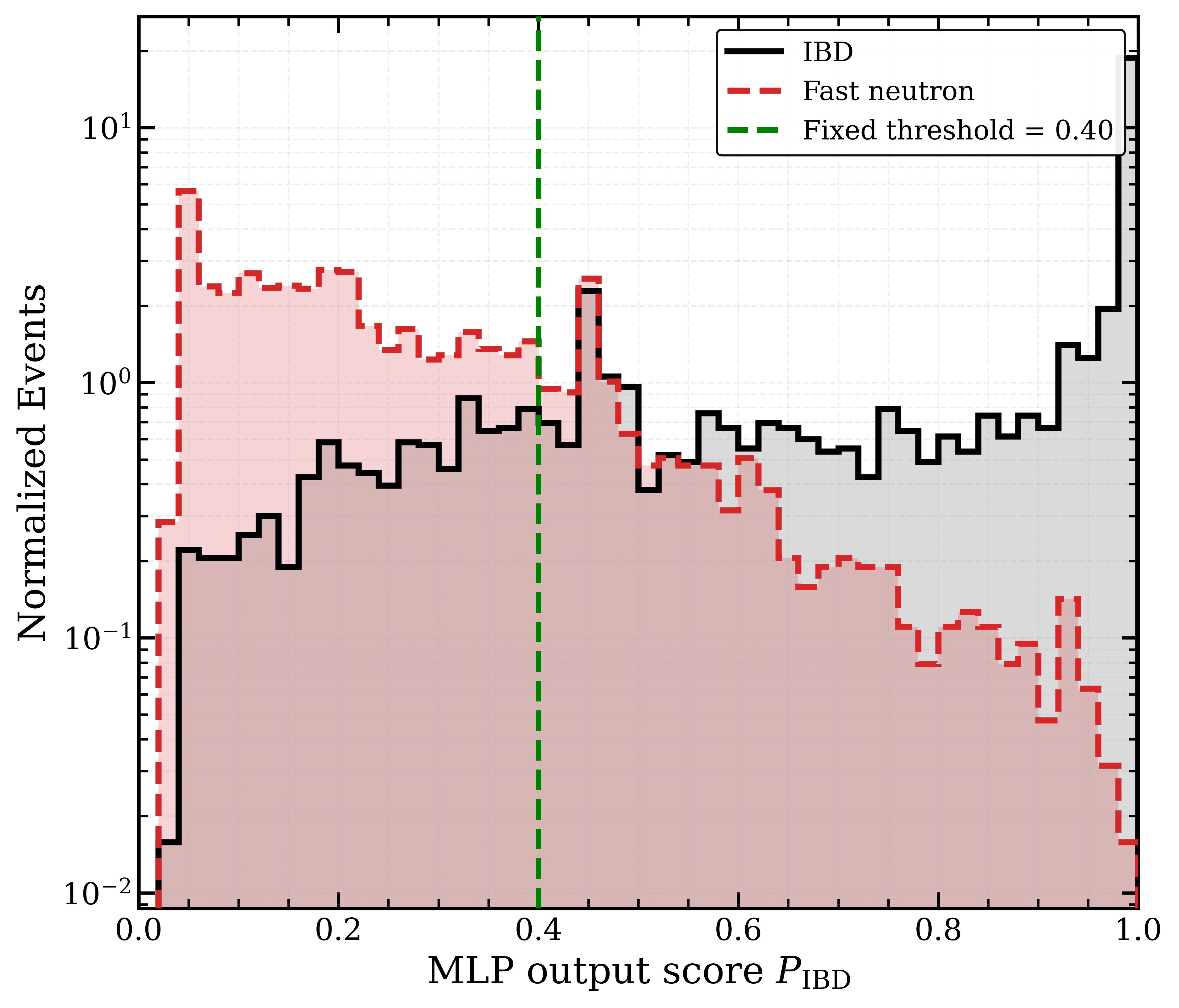}
\caption{
Normalized MLP classifier score distributions for simulated IBD signal and fast neutron background events evaluated using the independent testing sample. IBD events preferentially populate the high score region, whereas fast neutron events are concentrated at lower scores. The separation reflects the discrimination provided by the reconstructed prompt event topology. The vertical line indicates the frozen validation-derived threshold at $\mathrm{P_{IBD}}=0.40$, determined from the validation sample.}
\label{fig:mlp_score}
\end{figure*}

The MLP output, $\mathrm{P_{IBD}}$, provides a continuous discriminator between the prompt topologies associated with IBD positrons and fast neutron-induced proton recoils. Figure~\ref{fig:mlp_score} shows the normalized MLP score distributions for the simulated IBD signal and fast neutron background evaluated using the independent testing sample. IBD events preferentially populate the high-score region, whereas fast neutron events are concentrated toward lower classifier scores. The observed separation demonstrates that the combination of the three topology sensitive observables, $\mathrm{R_{WE}}$, $\mathrm{E_{D}}$, and $\mathrm{E_{max}^{hit}}$, provides substantial discrimination between the two event classes.

The separation has a direct physical interpretation. In an IBD interaction, the positron deposits its kinetic energy predominantly through electromagnetic interactions and subsequently annihilates into two $\mathrm{511~\mathrm{keV}}$ photons. Compton scattering and energy deposition by these photons can distribute the prompt energy among neighbouring scintillator bars, producing characteristic correlations between the total deposited energy, the highest energy bars, and the event multiplicity across the segmented detector. Fast neutron interactions, in contrast, are predominantly characterized by proton recoils, for which recoil protons have short ranges and comparatively large stopping power, producing more localized and irregular energy deposition patterns. The MLP combines these complementary features and maps their multidimensional correlations onto a single classifier score representing the compatibility of the reconstructed prompt event with the IBD class.

Events in the intermediate-score region correspond to prompt topologies for which the two interaction classes are less distinct. Such overlap is expected from fluctuations in the underlying energy deposition processes and from the finite detector resolution. The residual overlap therefore reflects both intrinsic interaction variability and detector response smearing.

Table~\ref{tab:mlp_performance} summarizes the final MLP operating point on the independent balanced testing sample. Quantities that depend on the class composition, such as purity and $S/B$, refer to the simulated $1:1$ signal-background reference sample and should therefore be interpreted as benchmark classification metrics rather than expected reactor-ON values.

\begin{table}[t]
\centering
\caption{Performance of the final MLP selection on the independent balanced testing sample at the fixed threshold $\mathrm{P_{IBD}>}0.40$. Purity and $S/B$ depend on the $1:1$ signal-background composition of the reference sample and are therefore benchmark classification metrics rather than expected experimental reactor-ON values.}
\label{tab:mlp_performance}
\begin{tabular}{lc}
\hline
Quantity & Value \\
\hline
IBD MLP classification efficiency & $(83.4\pm 0.7)\%$ \\
Fast neutron rejection & $(77.3\pm 0.7)\%$ \\
Signal purity, $f_{IBD}$ & 78.61\% \\
$\mathrm{F_1}$ score & 0.809 \\
Signal to background ratio, $S/B$ & 3.67 \\
\hline
\end{tabular}
\end{table}

The score distribution and the threshold-dependent performance together demonstrate that the MLP converts the multidimensional prompt-event topology into an effective one-dimensional discriminating variable. This allows a substantial reduction of the correlated fast neutron component while retaining most of the IBD signal candidates, providing an improved starting point for the subsequent reactor ${\overline{\ensuremath{\nu}}}_{e}$ analysis.

\subsection{Robustness against the AmBe fast neutron energy spectrum}
\label{ambe}

The MLP is trained using fast neutrons generated with a uniform kinetic energy distribution between $2$ and $12~\mathrm{MeV}$. This choice provides broad coverage of neutron-induced proton recoil topologies, but it does not represent a unique experimental fast neutron spectrum. To test the dependence of the classifier response on this assumption, an independent application sample containing fast neutrons generated according to an AmBe neutron spectrum is processed with the same detector reconstruction, input observables, feature standardization, and trained MLP~\cite{ambe}. Importantly, the operating point is not re-optimized: the validation-derived requirement
\begin{equation}
P_{\rm IBD}>0.40
\end{equation}
is retained unchanged.

For the independent AmBe application sample, the classifier gives an AUC of $0.9301$. At the fixed operating point, the MLP classification efficiency for the IBD population contained in this application sample is

\begin{equation}
\epsilon_{IBD}^{MLP}
=
(76.83\pm 0.07)\% ,
\end{equation}

while the AmBe fast neutron rejection is

\begin{equation}
R_{FN}^{AmBe}
=
(89.71\pm 0.10)\% ,
\end{equation}

corresponding to a residual fast neutron leakage of $10.29\%$. 

For comparison, the uniformly generated $2$-$12~\mathrm{MeV}$ fast neutron testing sample gives a rejection of $(77.3\pm 0.7)\%$ at the same fixed threshold. The change relative to the uniformly generated neutron sample demonstrates that the quantitative background acceptance depends on the incident neutron spectrum. The direction of this change is not interpreted as a universal improvement, rather, it reflects the different recoil-energy and topology distributions sampled by the two neutron spectra. 

Nevertheless, substantial fast neutron rejection is retained for the independent AmBe population without retraining or threshold adjustment. The discrimination is therefore not tied exclusively to the uniform neutron spectrum adopted during training, although its quantitative background acceptance is spectrum dependent. This observation is important for application to reactor data, where reactor-OFF data or dedicated neutron control samples should be used to constrain the residual fast neutron spectrum and normalization after the fixed MLP selection.

\subsection{IBD prompt energy spectrum after MLP selection}
\label{sec:spectrum}

\begin{figure}[t]
\centering
\includegraphics[width=1.0\textwidth]{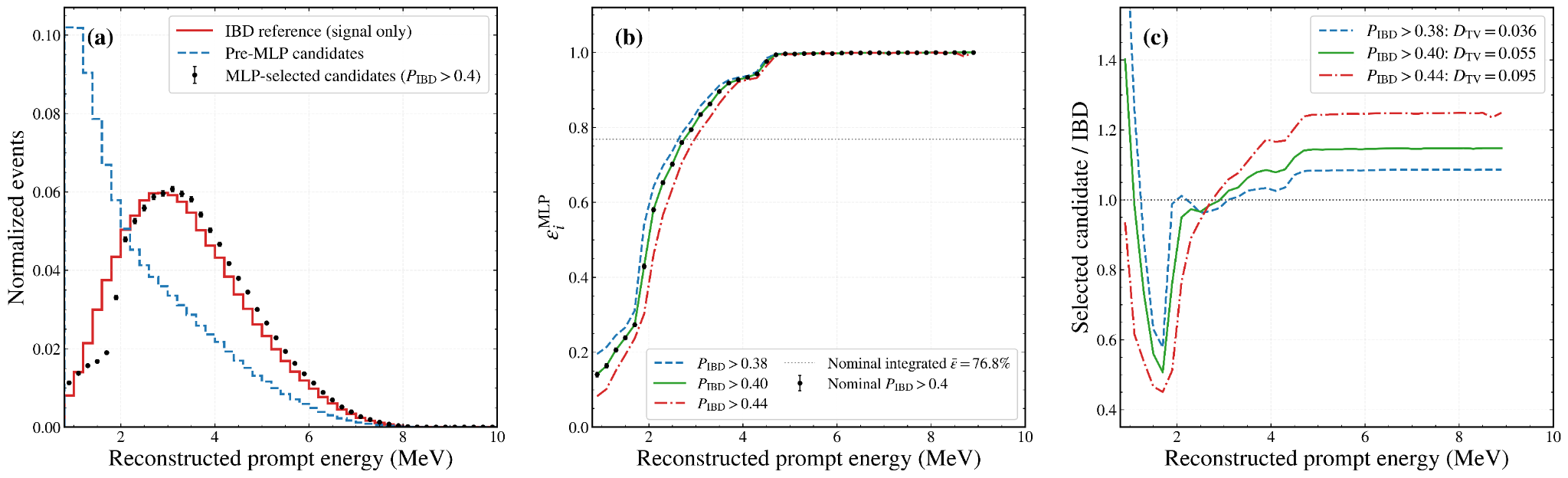}
\caption{Reconstructed prompt energy spectra for the controlled pre-MLP $S/B=1:1$ spectral benchmark. (a) Unit-normalized signal-only reconstructed IBD reference, mixed IBD + fast neutron candidate spectrum before the MLP requirement, and the same candidate population after the fixed selection $\mathrm{P_{IBD}>}0.40$. The selection increases the expected signal-to-background ratio from $1.00$ to $7.47$ and reduces the total-variation distance of the mixed candidate spectrum from the IBD reference from $0.343$ to $0.055$, corresponding to an $84.0\%$ reduction of the net candidate spectrum deformation. (b) Energy dependent IBD acceptance for the nominal selection and the diagnostic thresholds $\mathrm{P_{IBD}>}0.38$, $\mathrm{P_{IBD}>}0.40$, and $\mathrm{P_{IBD}>}0.44$. Black points show the nominal acceptance, and the horizontal dotted line indicates its integrated value of $76.8\%$. Error bars in panels (a) and (b) indicate finite-MC statistical uncertainties. (c) Ratios of the separately unit-normalized selected candidate spectra to the pre-MLP signal-only IBD reference for the three diagnostic thresholds. Their total-variation distances are $0.036$, $0.055$, and $0.095$, respectively. The horizontal dotted line denotes unity. The threshold comparison is diagnostic only, the nominal operating point remains fixed by the independent validation sample.}
\label{fig:ibd_spectrum}
\end{figure}

The impact of the topology-based selection on the reconstructed prompt energy spectrum is examined in Fig.~\ref{fig:ibd_spectrum}. A controlled pre-MLP composition $S/B=1:1$ is adopted to quantify how fast neutron contamination modifies the candidate spectrum and how this composition changes after application of the fixed classifier requirement. This controlled normalization is used only as a spectral shape benchmark. The more background dominated pseudo-experiment used for the likelihood-based signal extraction closure study is introduced separately in Sec.~\ref{sec:likelihood}.

Figure~\ref{fig:ibd_spectrum}(a) compares the signal-only reconstructed IBD reference with the mixed candidate distributions before and after the MLP requirement. For this spectral-shape study, the relative normalizations of the two components are imposed such that the pre-MLP candidate population has $S/B=1:1$. This controlled composition is used only to quantify the fast neutron-induced spectral deformation and its modification by the classifier, it is not intended to represent the physical reactor-ON signal-to-background ratio. The expected signal-to-background ratio therefore improves from
\begin{equation}
(S/B)_{\rm pre}=1.00
\quad\longrightarrow\quad
(S/B)_{\rm post}=7.47 ,
\end{equation}
and the IBD fraction of the candidate sample increases from $50.0\%$ to $88.19\%$.

The corresponding reduction in candidate spectrum deformation relative to pre-MLP candidate spectrum is quantified using the total variation distance,
\begin{equation}
D_{TV}(p,q) = \frac{1}{2}\sum_i\left|p_i-q_i\right| ,
\label{eq:dtv}
\end{equation}
where $\mathrm{p_{i}}$ and $\mathrm{q_{i}}$ denote unit-normalized reconstructed prompt energy spectra. Before application of the MLP requirement, the mixed candidate population gives
\begin{equation}
D_{TV}^{pre}=0.343
\end{equation}
relative to the signal-only IBD reference. After the fixed MLP selection, the corresponding mixed candidate spectrum gives
\begin{equation}
D_{TV}^{MLP}=0.055 .
\end{equation}
The net candidate spectrum deformation relative to the pre-MLP IBD reference is therefore reduced by
\begin{equation}
1-\frac{D_{\rm TV}^{\rm MLP}}{D_{\rm TV}^{\rm pre}}=84.0\% .
\end{equation}

This reduction should not, however, be interpreted as evidence that the classifier leaves the genuine IBD spectral shape unchanged. To isolate the effect of the MLP acceptance on the signal itself, the signal-only IBD population passing the same $\mathrm{P_{IBD}>}0.40$ requirement is compared with the pre-MLP IBD reference. The resulting value,
\begin{equation}
D_{TV}^{IBD,sel}=0.123,
\end{equation}
demonstrates a non-negligible energy dependence of the IBD selection efficiency. Since
\begin{equation}
D_{TV}^{MLP} < D_{TV}^{IBD,sel},
\end{equation}
the close agreement of the full MLP-selected candidate spectrum with the pre-MLP IBD reference arises from the combined effect of the energy dependent signal acceptance and the residual fast neutron component. In the normalized mixed spectrum these effects partially compensate each other. Consequently, the small value of $\mathrm{D_{TV}^{MLP}}$ alone should not be interpreted as direct spectral-shape preservation of the selected IBD population.

Figure~\ref{fig:ibd_spectrum}(b) directly displays the energy dependent IBD acceptance. For reconstructed prompt energy bin $i$ and classifier threshold $t$, the acceptance is defined as
\begin{equation}
\epsilon_i^{\rm MLP}(t)
=
\frac{
\sum_{k\in{\rm IBD},\,i}
\mathbf{1}(P_{{\rm IBD},k}>t)
}{
N_{{\rm IBD},i}
},
\label{eq:mlp_acceptance}
\end{equation}
where the sums run over IBD events satisfying the common pre-selection, and $\mathbf{1}$ is the selection indicator. At the nominal threshold $t=0.40$, the acceptance is strongly suppressed at low reconstructed prompt energy and approaches unity above $\sim$ $\mathrm{5~{MeV}}$. The integrated acceptance is $\bar{\epsilon}^{MLP}=76.8\%$, shown by the horizontal dotted line. This integrated value summarizes the selected signal yield but does not replace the bin-dependent acceptance required for a spectral analysis.

Figure~\ref{fig:ibd_spectrum}(c) shows the ratios of the separately unit-normalized selected candidate spectra to the pre-MLP signal-only IBD reference for $t=0.38$, $0.40$, and $0.44$. The corresponding total variation distances are $0.036$, $0.055$, and $0.095$, respectively. These variations reflect changes in both the energy-dependent signal acceptance and the residual fast neutron contribution. Because each spectrum is normalized separately, departures of these ratios from unity describe spectral-shape differences rather than absolute selection efficiencies. In particular, a ratio above unity at high energy is compatible with an IBD acceptance approaching unity in panel (b).

The three thresholds are used only to diagnose the dependence of the selected spectrum on the classifier requirement.The smaller total variation distance obtained at $t=0.38$ is not used to redefine the nominal operating point:$\mathrm{P_{IBD}>0.40}$ remains the requirement determined from the independent validation sample. The ratios are descriptive shape comparisons, and no independent goodness-of-fit probability is assigned to these correlated pre- and post-selection spectra.

The spectral benchmark therefore demonstrates two complementary features of the topology-based selection. First, it strongly suppresses the fast neutron contribution, increasing the IBD fraction from $50.0\%$ to $88.19\%$ for the controlled $S/B=1:1$ mixture. Second, the classifier introduces a measurable energy dependence in the IBD acceptance. A reactor ${\overline{\ensuremath{\nu}}}_{e}$ spectral analysis must therefore propagate this bin-dependent response rather than assuming that the selected signal spectrum is an unbiased representation of the pre-selection IBD shape. 

Accordingly, in Sec.~\ref{sec:likelihood} the detector-level selected spectrum is modeled as the sum of the IBD and residual fast neutron components. The bin-dependent IBD efficiency is incorporated explicitly in the signal expectation, while the residual fast neutron normalization is constrained simultaneously in the fit. The likelihood analysis then tests whether the underlying reconstructed IBD spectrum and overall signal normalization can be recovered after topology-based background suppression.

\section{Likelihood-based reactor ${\overline{\ensuremath{\nu}}}_{e}$ signal extraction}
\label{sec:likelihood}

The topology-based MLP selection is intended not only to increase the IBD content of the candidate sample, but also to test whether the IBD normalization can be recovered with controlled bias of the reactor ${\overline{\ensuremath{\nu}}}_{e}$ signal in the presence of residual fast neutron backgrounds. Figure~\ref{fig:ibd_spectrum} demonstrates that the MLP substantially suppresses the fast neutron-induced deformation of the prompt energy spectrum, while the remaining difference relative to the pre-MLP IBD reference contains a non-negligible contribution from the energy-dependent signal acceptance. A spectral likelihood analysis is therefore used to test whether the IBD normalization and reconstructed prompt energy spectrum can be recovered after explicitly accounting for the residual fast neutron component and the energy dependent MLP efficiency.

The likelihood study is performed as a controlled MC pseudo experiment. The available reconstructed IBD and AmBe fast neutron samples are divided into statistically independent template and injection subsets. The template subset is used to construct the likelihood model, whereas the independent injection subset defines the expectation from which Poisson-fluctuated pseudo data are generated. This separation prevents an artificially exact closure resulting from using the same histogram for both generation and fitting.

The raw IBD to fast neutron ratio of the simulated samples is not used as the physical sample composition. Instead, the relative signal and background normalizations are imposed explicitly at the pseudo experiment level. The nominal benchmark corresponds to a pre-MLP ratio $\mathrm{S/B=1:2}$ before the MLP requirement. A more background dominated $\mathrm{S/B=1:4}$ scenario is used separately as a robustness stress test.

\subsection{Likelihood model}

For reconstructed prompt energy bin $i$, the nominal detector-level expectation before accounting for finite MC-template statistics is written as
\begin{equation}
\mu_i^{0} = \alpha
\left(1+\sigma_{\epsilon}\eta_{\epsilon}\right)
N_{S}^{\rm pre}\,
s_i^{\rm pre}\,
\epsilon_i
+
\beta\,
N_{B}^{\rm pre}\,
f_{\rm FN}^{\rm MLP}\,
b_i^{\rm MLP},
\label{eq:likelihood_expectation}
\end{equation}
where $\mathrm{s_i^{pre}}$ is the unit normalized reconstructed IBD prompt energy spectrum before the MLP requirement, $\mathrm{\epsilon_{i}}$ is the bin-dependent IBD selection efficiency of the fixed requirement $\mathrm{P_{IBD}>0.40}$, and $\mathrm{b_i^{MLP}}$ is the unit normalized residual fast neutron spectrum after the same selection. The quantity $f_{FN}^{MLP}$ denotes the integrated fast neutron leakage fraction. The parameter $\alpha$ is the IBD signal strength parameter of interest, while $\beta$ scales the residual fast neutron normalization.

A correlated MLP efficiency scale uncertainty is introduced through the standard normal nuisance parameter $\mathrm{\eta_{\epsilon}}$, with $\mathrm{\sigma_{\epsilon}=0.05}$. The residual fast neutron normalization is constrained with a Gaussian width $\mathrm{\sigma_{B}=0.10}$ around $\beta=1$. These values are used as benchmark nuisance widths for the MC pseudo experiment.


Because statistically independent MC subsets are used for the injected and fitted spectra, the finite statistics of the likelihood templates must also be propagated. A Gaussian-constrained Barlow-Beeston-lite treatment is therefore introduced through a bin-wise multiplicative nuisance parameter $\gamma_i$~\cite{BarlowBeeston},
\begin{equation}
\mu_i = \gamma_i\,\mu_i^{0},
\end{equation}
with $\mathrm{\gamma_{i}}$ constrained by the statistical uncertainty of the corresponding MC template. 


The likelihood is then
\begin{equation}
\mathcal{L}
=
\left[
\prod_i
{\rm Pois}(n_i|\mu_i)\,
G(\gamma_i;1,\sigma_{{\rm MC},i})
\right]
G(\beta;1,\sigma_B)\,
G(\eta_{\epsilon};0,1).
\label{eq:likelihood_full}
\end{equation}
where $\mathrm{n_{i}}$ is the pseudo-data count in bin $i$ and $\mathrm{\sigma_{{MC},i}}$ represents the relative finite template uncertainty. The $\mathrm{\gamma_{i}}$ parameters are profiled analytically in each bin, while $\beta$ and $\mathrm{\eta_{\epsilon}}$ are profiled numerically. The signal and background template uncertainties are combined in quadrature to define $\sigma_{{MC},i}$ for the common Gaussian-constrained Barlow-Beeston-lite nuisance $\gamma_i$. The reconstructed prompt energy interval $\mathrm{0.8<E_{prompt}<10.0~\mathrm{MeV}}$ is divided into 46 uniform bins of width $0.20~\mathrm{MeV}$.


The IBD normalization is determined from the profile likelihood in $\alpha$. The 68\% interval is obtained from $\mathrm{\Delta(-2\ln\mathcal{L})=1}$. Since the likelihood contains constrained nuisance parameters and finite template uncertainties, the goodness of fit is evaluated using parametric pseudo experiments. The empirical goodness of fit probability is defined as the fraction of bootstrap pseudo experiments whose Poisson deviance exceeds that of the observed pseudo-data realization.

\subsection{Nominal IBD signal extraction}

\begin{figure}[t]
\centering
\includegraphics[width=0.95\textwidth]{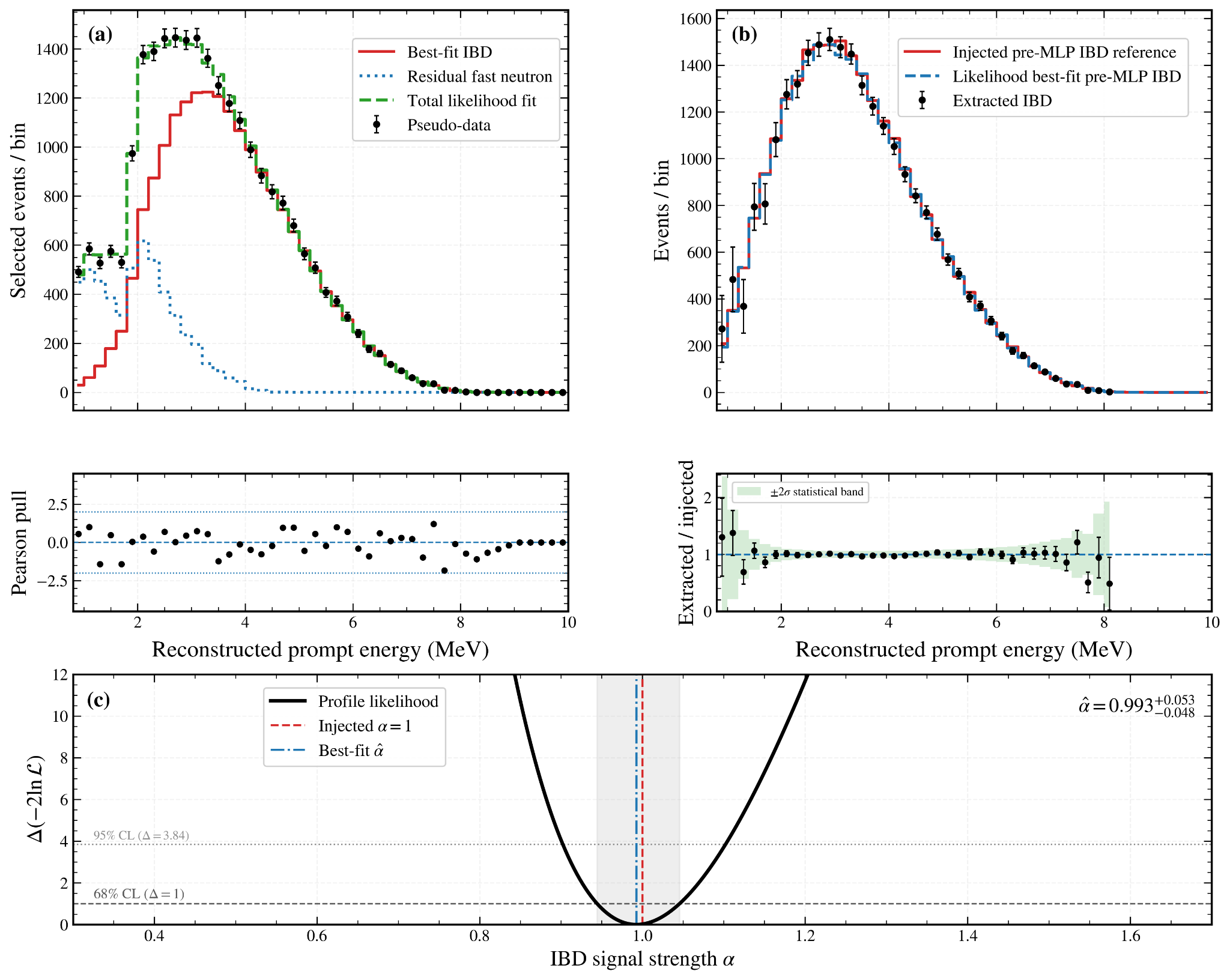}
\caption{
Likelihood-based reactor IBD signal extraction for the nominal pre-MLP $\mathrm{S/B=1:2}$ pseudo-experiment. (a) Detector level prompt energy spectrum after the fixed $\mathrm{P_{IBD}>}0.40$ requirement. The points show the Poisson fluctuated pseudo-data, while the curves show the best-fit IBD contribution, residual fast neutron background, and their sum. The lower panel shows the bin-wise pulls. (b) Reconstructed IBD spectrum obtained after subtraction of the fitted fast neutron contribution and correction for the energy-dependent MLP signal efficiency. The extracted spectrum is compared with the independently generated pre-MLP IBD reference, the lower panel shows their ratio. (c) Profile likelihood for the IBD signal-strength parameter $\alpha$. The vertical lines indicate the injected value $\alpha_{\rm inj}=1$ and the best-fit value, while the horizontal lines correspond to $\mathrm{\Delta(-2\ln\mathcal{L})=1}$ and $3.84$. Finite MC-template statistics are included using a Barlow-Beeston-lite treatment.}
\label{fig:mlp_likelihood_fit}
\end{figure}

Figure~\ref{fig:mlp_likelihood_fit} shows the result for the nominal pre-MLP composition $\mathrm{S/B=1:2}$. The imposed pre-MLP expectations are $\mathrm{N_{S}^{pre}=2.5\times10^{4}}$ IBD events and $\mathrm{N_{B}^{pre}= 5.0\times10^{4}}$ fast neutron events. After application of the fixed $\mathrm{P_{IBD}>}0.40$ requirement, the independent injection sample contains expected selected yields of $\mathrm{1.919\times10^{4}}$ IBD events and $\mathrm{5.153\times10^{3}}$ residual fast neutron events. The corresponding signal-to-background ratio changes from
\begin{equation}
(S/B)_{pre}=0.50
\quad\longrightarrow\quad
(S/B)_{post}=3.724,
\end{equation}
and the expected IBD fraction increases from $33.3\%$ before the MLP selection to $78.83\%$ afterwards.

The detector-level fit shown in Fig.~\ref{fig:mlp_likelihood_fit}(a) reproduces the selected prompt energy spectrum with
\begin{equation}
\hat{\alpha}
=
0.993^{+0.053}_{-0.048},
\qquad
\hat{\beta}=1.026 .
\end{equation}
The injected value $\mathrm{\alpha_{inj}=1}$ lies within the 68\% profile interval. The parametric bootstrap goodness of fit probability is
\begin{equation}
p_{emp}=0.434,
\end{equation}

indicating that the detector-level signal plus background model is compatible with the pseudo data realization.

The physical role of the likelihood is illustrated in Fig.~\ref{fig:mlp_likelihood_fit}(b). The fitted residual fast neutron component is subtracted from the selected candidate spectrum, and the result is corrected bin by bin for the MLP signal efficiency. The extracted pre-MLP IBD spectrum is compared with the statistically independent injected IBD reference. Using the propagated pseudo-data counting uncertainty gives the diagnostic closure statistic

\begin{equation}
\chi^2_{stat}/N_{dof}=25.95/34,
\end{equation}
with a corresponding diagnostic $p$-value of $0.84$. The RMS bin-by-bin deviation of the extracted to injected ratio is $16.0\%$. The largest deviations occur in the low-statistics edges of the spectrum, whereas the central prompt energy region follows the injected IBD distribution closely. These quantities are used as spectral-closure diagnostics rather than as the primary likelihood goodness-of-fit test.

Figure~\ref{fig:mlp_likelihood_fit}(c) shows the profile likelihood for the IBD signal strength. The minimum occurs close to the injected normalization, demonstrating that the energy dependent signal acceptance introduced by the MLP can be propagated through the likelihood with the injected value lying within the 68\% profile interval in the recovered integrated IBD yield.

\subsection{Uncertainty budget and pseudo-experiment validation}

The relative uncertainty on the extracted IBD normalization is evaluated by successively introducing the finite-template, fast neutron, and MLP efficiency nuisance terms. The resulting uncertainty decomposition for the nominal benchmark is summarized in Table~\ref{tab:likelihood_uncertainty}.

\begin{table}[t]
\centering
\caption{Relative 68\% profile-likelihood half-width on the IBD signal-strength parameter $\alpha$ for the nominal pre-MLP $S/B=1:2$ pseudo-experiment. Nuisance contributions are introduced successively, the 5\% MLP efficiency and 10\% fast neutron normalization widths are benchmark assumptions for this controlled MC study.}
\label{tab:likelihood_uncertainty}
\begin{tabular}{lc}
\hline
Likelihood configuration & Relative uncertainty on $\alpha$ \\
\hline
Statistical only
& $0.772\%$ \\
Statistics + finite MC-template statistics
& $0.883\%$ \\
Statistics + MC statistics + FN normalization
& $0.950\%$ \\
Full model including MLP-efficiency scale
& $5.102\%$ \\
\hline
\end{tabular}
\end{table}

Finite statistics of the independent MC templates increase the relative 68\% uncertainty from $0.772\%$ to $0.883\%$. Adding the assumed 10\% residual fast neutron normalization uncertainty increases it to $0.950\%$. The full uncertainty becomes $5.102\%$ after inclusion of the assumed 5\% correlated MLP-efficiency scale uncertainty. Under these benchmark nuisance assumptions, the uncertainty on the extracted IBD normalization is therefore dominated by knowledge of the MLP signal efficiency rather than by counting statistics or residual fast neutron normalization.

The stability of the estimator is further examined using ensembles of pseudo-experiments. With the nuisance truths held fixed at their nominal values, 999 of 1000 pseudo-experiments converge successfully and give

\begin{equation}
\left\langle\hat{\alpha}\right\rangle = 0.9951,
\qquad
{\rm RMS}(\hat{\alpha})=0.0085,
\end{equation}
corresponding to a mean relative bias of $-0.49\%$. When the MLP efficiency and fast neutron nuisance truths are varied according to their benchmark constraints, 1000 pseudo-experiments give
\begin{equation}
\left\langle\hat{\alpha}\right\rangle = 0.9954,
\qquad
{\rm RMS}(\hat{\alpha})=0.0508,
\end{equation}
with a mean relative bias of $-0.46\%$. The increase in the ensemble width to $\sim$ $5\%$ is consistent with the dominant correlated MLP-efficiency scale uncertainty obtained from the profile-likelihood decomposition.

\subsection{Robustness against increased fast neutron contamination}
\label{sec:likelihood_stress}

The stability of the likelihood-based IBD extraction is further examined under a more background-dominated condition. Starting from the nominal pre-MLP composition $S/B=1:2$, the expected fast neutron population is increased by a factor of two, giving a stress test composition $S/B=1:4$. The pre-MLP IBD expectation is kept fixed at $\mathrm{N_{S}^{pre}=2.5\times10^{4}}$ events, while the fast neutron expectation is increased from $5.0\times10^{4}$ to $1.0\times10^{5}$ events. All other ingredients of the analysis are unchanged: the same reconstructed prompt energy interval, fixed $\mathrm{P_{IBD}>}0.40$ requirement, independent template/injection MC split, and benchmark nuisance constraints are retained.

For the high-background configuration, the expected post-MLP yields are

\begin{equation}
N_{S}^{\rm sel}=1.919\times10^{4},
\qquad
N_{B}^{\rm sel}=1.031\times10^{4},
\end{equation}

so that the signal-to-background ratio changes from

\begin{equation}
(S/B)_{\rm pre}=0.25
\quad\longrightarrow\quad
(S/B)_{\rm post}=1.862 .
\end{equation}

The corresponding IBD fraction increases from $20.0\%$ before the MLP requirement to $65.06\%$ after the topology-based selection. For comparison, the nominal $S/B=1:2$ scenario gives $\mathrm{(S/B)_{post}=}3.724$ and an IBD fraction of $78.83\%$.

\begin{table}[t]
\centering
\caption{
Comparison of the likelihood-based IBD extraction for the nominal pre-MLP $S/B=1:2$ configuration and the $S/B=1:4$ high-background stress test. The injected IBD normalization, $\mathrm{P_{IBD}>}0.40$ operating point, reconstructed energy binning, detector response, and benchmark nuisance widths are unchanged between the two scenarios. Finite MC-template statistics are propagated using the Barlow-Beeston-lite treatment. The $S/B=1:4$ configuration is a robustness stress test and is not included as an additional independent systematic uncertainty.}
\label{tab:likelihood_sb_stress}

\begin{tabular}{lcc}
\hline
Quantity
& Nominal $S/B=1:2$
& Stress test $S/B=1:4$ \\
\hline

Post-MLP $S/B$
& $3.724$
& $1.862$ \\

Post-MLP IBD fraction
& $78.83\%$
& $65.06\%$ \\

$\hat{\alpha}$
& $0.993^{+0.053}_{-0.048}$
& $0.987^{+0.053}_{-0.048}$ \\

$\hat{\beta}$
& $1.026$
& $1.027$ \\

Poisson deviance/$N_{\rm dof}$ (diagnostic)
& $26.97/43$
& $19.61/43$ \\

Empirical GOF $p$-value
& $0.434$
& $0.613$ \\

Spectral-closure RMS
& $16.0\%$
& $16.6\%$ \\

Fixed-toy $\langle\hat{\alpha}\rangle$
& $0.9951$
& $0.9898$ \\

Fixed-toy RMS$(\hat{\alpha})$
& $0.0085$
& $0.0088$ \\

Fixed-toy mean bias
& $-0.49\%$
& $-1.02\%$ \\

Full relative 68\% uncertainty
& $5.102\%$
& $5.115\%$ \\

\hline
\end{tabular}
\end{table}

Applying the same likelihood model to the high-background pseudo-experiment gives

\begin{equation}
\hat{\alpha} = 0.987^{+0.053}_{-0.048},
\qquad
\hat{\beta}=1.027 ,
\end{equation}

for an injected signal strength $\alpha_{inj}=1$. The injected normalization therefore remains well within the 68\% profile-likelihood interval,

\begin{equation}
0.939 < \alpha < 1.040 .
\end{equation}

The preferred parametric bootstrap goodness-of-fit probability is

\begin{equation}
p_{emp}=0.613 ,
\end{equation}

showing that the likelihood model remains compatible with the pseudo-data after doubling the pre-MLP fast neutron contribution. The corresponding Poisson deviance is $19.61$ for a diagnostic $\mathrm{N_{dof}=43}$.

After subtraction of the fitted residual fast neutron component and correction for the energy-dependent MLP signal efficiency, the reconstructed pre-MLP IBD spectrum remains consistent with the independently injected IBD reference. The spectral-closure diagnostic is

\begin{equation}
\chi^2_{stat}/N_{dof} = 25.01/34 ,
\end{equation}

with a diagnostic $p$-value of $0.869$ and an RMS fractional bin-by-bin deviation of $16.6\%$. This is close to the $16.0\%$ closure RMS obtained in the nominal configuration, indicating that the spectral reconstruction is not strongly degraded by the increased fast neutron normalization.

The profile likelihood uncertainty on the IBD normalization is also only weakly affected. The full relative 68\% half-width changes from $5.102\%$ in the nominal scenario to $5.115\%$ in the high-background scenario. Thus, under the benchmark nuisance assumptions adopted here, the total uncertainty remains dominated by the correlated 5\% MLP-efficiency scale rather than by the increased fast neutron statistics.

An ensemble of fixed-nuisance pseudo-experiments provides an additional robustness test. For the $S/B=1:4$ configuration, 999 of 1000 fits converge successfully and give

\begin{equation}
\left\langle\hat{\alpha}\right\rangle = 0.9898,
\qquad
{\rm RMS}(\hat{\alpha})=0.0088 ,
\end{equation}

corresponding to a mean relative bias of $-1.02\%$. The corresponding nominal values are $\left\langle\hat{\alpha}\right\rangle=0.9951$, $\mathrm{{RMS}(\hat{\alpha})=}0.0085$, and a mean bias of $-0.49\%$. The increased fast neutron contamination therefore introduces a small additional downward bias in this controlled pseudo-experiment study, but the effect remains at the percent level and is substantially smaller than the full $\sim$ 5\% uncertainty on the extracted signal normalization.

The $S/B=1:4$ configuration is used only as a robustness stress test and is not treated as an additional independent systematic uncertainty. Overall, the comparison demonstrates that the likelihood framework continues to recover the injected IBD normalization and prompt energy spectrum when the pre-MLP correlated fast neutron contamination is doubled, provided that the residual background component and the energy dependent MLP acceptance are explicitly included in the fit.

\section{Implications for a sterile-neutrino search with ISMRAN}
\label{sec:sterile}

The preceding sections demonstrate that the MLP suppresses correlated fast neutron backgrounds and that the associated energy-dependent IBD acceptance can be propagated through a spectral likelihood. This capability is particularly relevant for very short baseline reactor ${\overline{\ensuremath{\nu}}}_{e}$ measurements, where one of the principal physics motivations is the search for oscillations involving an additional sterile neutrino state. In a minimal $3+1$ framework, the ${\overline{\ensuremath{\nu}}}_{e}$ survival probability at short baseline can be approximated by

\begin{equation}
P_{ee}(E_\nu,L) = 1-\sin^{2}2\theta_{14}
\sin^{2}
\left(
1.27\,
\frac{\Delta m^{2}_{41}[\mathrm{eV}^{2}]\,L[\mathrm{m}]}
{E_{\nu}[\mathrm{MeV}]}
\right),
\label{eq:sterile_survival}
\end{equation}

where $\mathrm{\Delta m_{41}^{2}}$ and $\mathrm{\theta_{14}}$ denote the additional mass-squared splitting and mixing angle, respectively. The resulting oscillation signature is therefore encoded not only in the total IBD rate but, more importantly for a segmented short-baseline experiment, in the baseline-dependent distortion of the reconstructed prompt energy spectrum. The number of events in the $\mathrm{i^{th}}$ reconstructed prompt energy bin after incorporating the detector resolution is given as,

\begin{equation}
S_{dij}^{\rm osc}
=
\mathcal{N}_{d}
\int dE_{\nu}\,
\Phi(E_{\nu})\,
\sigma_{\rm IBD}(E_{\nu})\,
P_{ee}(L_{d},E_{\nu})\,
R_{dij}(E_{\nu}) \, ,
\label{eq:sterile_signal_expectation}
\end{equation}

Here, $\mathcal{N}_{d}$ contains the exposure and overall detector normalization at baseline $d$, $\mathrm{\Phi(E_{\nu})}$ is the reactor ${\overline{\ensuremath{\nu}}}_{e}$ flux, $\mathrm{\sigma_{IBD}(E_{\nu})}$ is the IBD cross section, $\mathrm{P_{ee}(L_{d},E_{\nu})}$ is the ${\overline{\ensuremath{\nu}}}_{e}$ survival probability, and $\mathrm{R_{dij}(E_{\nu})}$ denotes the detector response mapping the true ${\overline{\ensuremath{\nu}}}_{e}$ energy into reconstructed prompt energy bin $i$ and MLP-score bin $j$ at baseline $d$.

The ISMRAN detector provides a useful configuration for such a search because measurements can be performed at multiple baselines from the Dhruva research reactor. The present sensitivity study considers the experimentally relevant baselines of 13~m and 18~m with two years of exposure at each position. ISMRAN has already accumulated approximately these exposures at both sites, so the configuration considered here corresponds directly to the existing experimental programme rather than to a purely hypothetical detector arrangement. The two baseline configuration provides sensitivity to the relative $L/E_{\nu}$ dependence of the reconstructed spectrum and reduces the dependence of the oscillation search on the absolute reactor flux normalization. The principal detector, reactor, exposure, background, and likelihood assumptions used for the sterile-neutrino sensitivity projection are summarized in Table~\ref{tab:sterile_assumptions}. These inputs are held fixed when comparing the energy-only and topology-assisted likelihoods, so that the difference between the two sensitivities arises from the additional information carried by the MLP classifier score.

\begin{table}[t]
\centering
\caption{
Nominal assumptions used for the ISMRAN sterile-neutrino sensitivity projection. The energy-only and $\mathrm{E_{prompt}\times P_{IBD}}$ analyses use the same reactor, detector, exposure, event population, fast neutron normalization, and nuisance constraints. The nominal fast neutron normalization corresponds to a controlled pre-MLP $S/B=1:2$ benchmark and is not interpreted as a measured reactor-ON background rate.}
\label{tab:sterile_assumptions}

\begin{tabular}{p{0.40\textwidth} p{0.48\textwidth}}
\hline
Parameters & Nominal value \\
\hline

Reactor thermal power & $100~\mathrm{MW_{th}}$ \\

Dhruva core model & Cylindrical core, $R=1.50\mathrm{m}$ and $H=3.03\mathrm{m}$ \\

Fission fractions & $^{235}\mathrm{U}:0.58$,$^{239}\mathrm{Pu}:0.30$,$^{241}\mathrm{Pu}:0.05$,$^{238}\mathrm{U}:0.07$ \\

Detector mass & $1.0~\mathrm{ton}$ plastic scintillator \\

IBD detection efficiency & $37.7\%$ \\


Reactor duty cycle & $70\%$ \\

Oscillation baselines & $L_{near}=13~\mathrm{m}$, $L_{far}=18~\mathrm{m}$ \\

Exposure & $2$ yr at each baseline \\

Prompt-event reference population
& $0.8<E_{\rm prompt}<10.0~\mathrm{MeV}$,
$2\leq N_{\rm bars}^{\rm prompt}\leq4$ \\

Reconstructed-energy binning
& $23$ uniform bins over
$0.8<E_{\rm prompt}<10.0~\mathrm{MeV}$ \\

MLP-score binning & $P_{\rm IBD}\in[0,0.2),[0.2,0.4),[0.4,0.6),[0.6,0.8),[0.8,1.0]$ \\

Nominal pre-MLP background level & Combined $S/B=1:2$ \\

MLP response in the nominal 2D likelihood & Fixed MC-derived $E_{ prompt}\times P_{IBD}$ response \\

MLP acceptance uncertainty & $5\%$ correlated scale used only for the $P_{IBD}>0.40$ acceptance-correction diagnostic \\

\hline
\end{tabular}
\end{table}

\subsection{Energy-only and topology-assisted oscillation likelihoods}

\begin{figure}[t]
\centering
\includegraphics[width=0.75\textwidth]{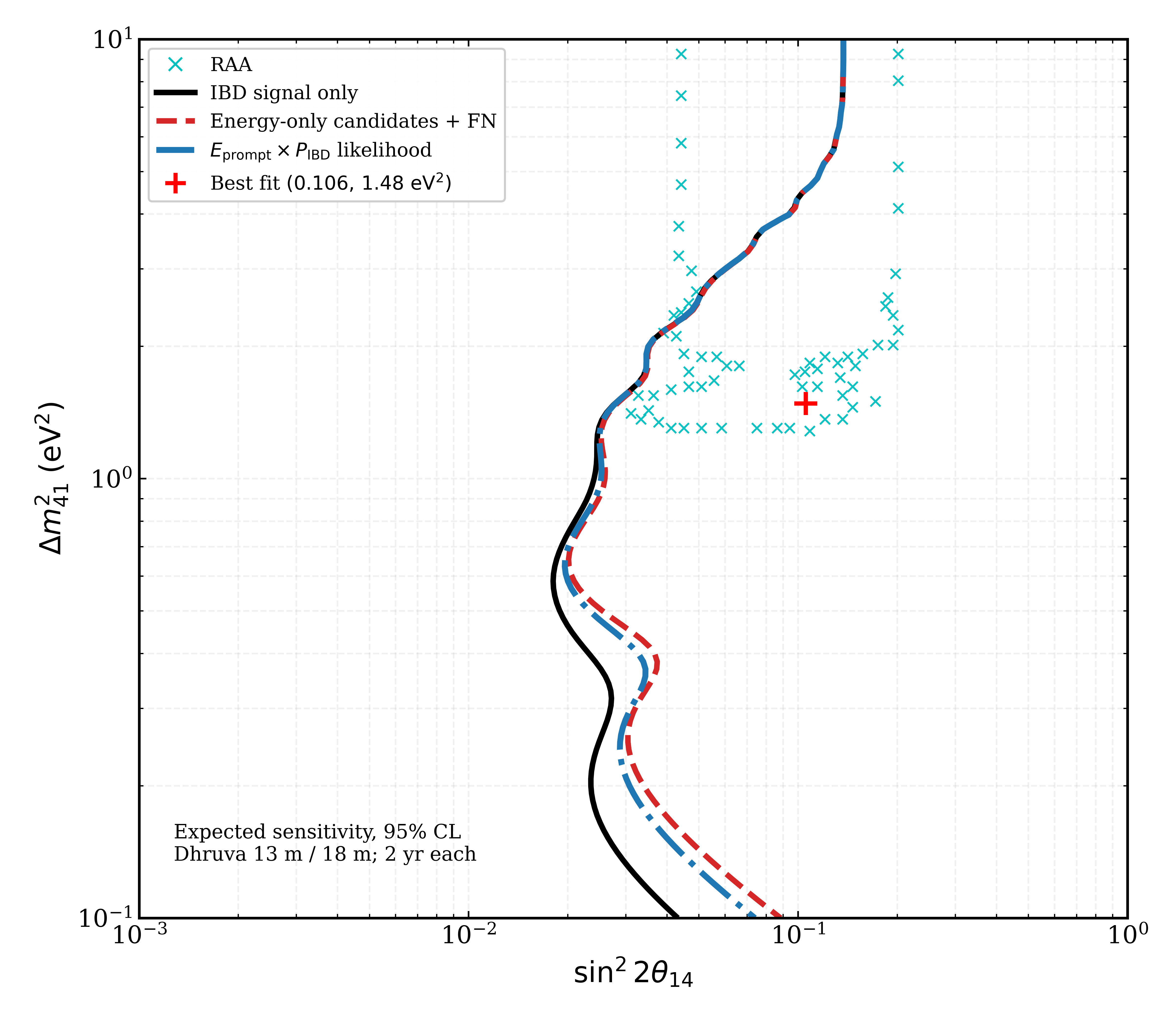}
\caption{Expected sterile-neutrino sensitivity for the ISMRAN 13~m and 18~m configuration with two years of exposure at each baseline. The cyan crosses reproduce the historical RAA reference points shown in Fig. 6 of Ref.~\cite{Behera2020Sterile}, based on the RAA interpretation of Ref.~\cite{Mention2011RAA}. The black curve shows the ideal IBD signal-only sensitivity, the red dashed curve includes the fast neutron contribution using reconstructed prompt energy alone, and the blue dash-dotted curve is obtained from the joint $\mathrm{E_{prompt}\times P_{IBD}}$ likelihood using the identical event population. The red marker denotes the reference best-fit point displayed in the published comparison. The nominal projection assumes a combined pre-MLP $S/B=1:2$.}


\label{fig:sterile_sensitivity}
\end{figure}

The sterile neutrino sensitivity is evaluated with the same detector response and fast neutron framework developed in the preceding sections. For the nominal study, a combined pre-MLP signal to background ratio of $S/B=1:2$ is imposed as a controlled fast neutron benchmark. 


For baseline $d$ and reconstructed prompt energy bin $i$, the expectation for the energy-only analysis is

\begin{equation}
\mu_{di}^{E}
=
\alpha\,
S_{di}^{\rm osc}
\left(
\sin^{2}2\theta_{14},
\Delta m_{41}^{2}
\right)
+
\beta_{d}\,
B_{di}^{\rm FN},
\label{eq:sterile_energy_likelihood}
\end{equation}

where $\mathrm{S_{di}^{osc}}$ is the oscillated IBD expectation after detector response, $\mathrm{B_{di}^{FN}}$ is the reconstructed fast neutron template, $\alpha$ is a common IBD normalization parameter, and $\mathrm{\beta_{d}}$ scales the fast neutron contribution independently at each detector position.

The topology-assisted analysis retains the MLP classifier score as an additional event observable. For MLP-score region $j$, the expectation becomes

\begin{equation}
\mu_{dij}^{E,P}
=
\alpha\,
S_{dij}^{\rm osc}
\left(
\sin^{2}2\theta_{14},
\Delta m_{41}^{2}
\right)
+
\beta_{d}\,
B_{dij}^{\rm FN}.
\label{eq:sterile_2d_likelihood}
\end{equation}

The final analysis uses five fixed score intervals, $0<P_{\rm IBD}<0.2$, $0.2<P_{\rm IBD}<0.4$, $0.4<P_{\rm IBD}<0.6$, $0.6<P_{\rm IBD}<0.8$, and $0.8<P_{\rm IBD}<1.0$. The reconstructed prompt energy interval $0.8<E_{\rm prompt}<10.0~\mathrm{MeV}$ is divided into 23 bins. Importantly,

\begin{equation}
S_{di}^{\rm osc}
=
\sum_j S_{dij}^{\rm osc},
\qquad
B_{di}^{\rm FN}
=
\sum_j B_{dij}^{\rm FN},
\label{eq:sterile_marginalization}
\end{equation}

so that the energy-only likelihood is exactly the classifier-score marginalization of the two-dimensional likelihood. The two analyses therefore contain the same events, the only difference is whether the additional topology information encoded in $\mathrm{P_{IBD}}$ is retained.

For either analysis, the binned Poisson likelihood can be written in deviance form as

\begin{align}
-2\ln\mathcal{L}
=&\;
2\sum_k
\left[
\mu_k-n_k+n_k\ln\left(\frac{n_k}{\mu_k}\right)
\right]
\nonumber\\
&+
\left(\frac{\alpha-1}{\sigma_{\alpha}}\right)^2
+
\left(\frac{\beta_{13}-1}{\sigma_{B,13}}\right)^2
+
\left(\frac{\beta_{18}-1}{\sigma_{B,18}}\right)^2 ,
\label{eq:sterile_likelihood}
\end{align}

where $k$ runs over $(d,i)$ for the energy-only analysis and over $(d,i,j)$ for the topology-assisted analysis. The sensitivity calculation includes a 3\% correlated IBD normalization uncertainty and independent fast neutron normalization uncertainties of 10\% and 6\% at 13 and 18~m, respectively. The nominal $\mathrm{E_{prompt}\times P_{IBD}}$ signal and background shapes are taken from the MC response templates and are kept fixed in the present expected sensitivity study.The oscillation parameters are scanned while the nuisance parameters are profiled. The expected 95\% C.L. sensitivity in the $(\sin^{2}2\theta_{14},\Delta m_{41}^{2})$ plane is defined by $\Delta\chi^{2}=5.99$, corresponding to two oscillation parameters under the Wilks approximation~\cite{Agostini2020SterileStatistics}. The resulting contours represent Asimov expected sensitivities, an exclusion derived from experimental data would require validation of the frequentist coverage using pseudo experiments~\cite{Cowan2011Asimov}.


This construction is physically different from applying the MLP as a hard event-selection requirement. No event is discarded by the classifier in the sterile-neutrino fit. Instead, low $\mathrm{P_{IBD}}$ regions provide fast neutron enriched control populations, whereas high $\mathrm{P_{IBD}}$ regions are increasingly enriched in IBD events. The classifier score therefore acts as an additional experimental observable with which the likelihood can constrain the correlated background while retaining the complete prompt event population for the oscillation measurement.

\subsection{Expected light sterile-neutrino sensitivity and physics interpretation}
\label{sec:sterile_results}

Figure~\ref{fig:sterile_sensitivity} shows the expected 95\% C.L. sterile-neutrino sensitivity for the ISMRAN 13~m and 18~m configuration. The historical RAA reference region is included for comparison~\cite{Mention2011RAA,Behera2020Sterile}. The black curve corresponds to an ideal IBD-only sample, the red dashed curve includes the fast neutron component but uses reconstructed prompt energy alone, and the blue dash-dotted curve retains the joint $\mathrm{E_{prompt}\times P_{IBD}}$ information. 

For the nominal combined pre-MLP $S/B=1:2$ benchmark, fast neutron contamination weakens the expected sensitivity relative to the ideal IBD-only case. Retaining the MLP-score dimension moves the sensitivity toward the ideal limit, demonstrating that part of the information lost through correlated background contamination can be recovered by using the reconstructed event topology. The nominal projection has sensitivity to a substantial part of the RAA-motivated parameter space, including the vicinity of the reference best-fit point shown in the published comparison. These curves are expected sensitivities and should not be interpreted as an experimental exclusion from the accumulated ISMRAN data.

The gain obtained from retaining the classifier topology information is quantified as

\begin{equation}
\mathcal{I}_{\rm MLP}(\Delta m_{41}^{2})
=
100\,
\frac{
s_{95}^{E}(\Delta m_{41}^{2})
-
s_{95}^{E,P}(\Delta m_{41}^{2})
}{
s_{95}^{E}(\Delta m_{41}^{2})
},
\label{eq:imlp}
\end{equation}

where $s_{95}^{E}$ and $s_{95}^{E,P}$ denote the 95\% C.L. limits on $\sin^{2}2\theta_{14}$ obtained from the energy-only and joint $\mathrm{E_{prompt}\times P_{IBD}}$ likelihoods, respectively. Positive values of $\mathcal{I}_{MLP}$ therefore correspond directly to improved sterile-neutrino mixing sensitivity from the event-topology information.

For the nominal $S/B=1:2$ benchmark, the logarithmically averaged improvement in the interval $0.1\leq\Delta m_{41}^{2}\leq1.0~\mathrm{eV}^{2}$ is

\begin{equation}
\left\langle
\mathcal{I}_{\rm MLP}
\right\rangle_{\log}
=
\frac{
\displaystyle
\int_{0.1}^{1.0}
\mathcal{I}_{\rm MLP}(\Delta m_{41}^{2})
\,d\ln\Delta m_{41}^{2}
}{
\ln(1.0/0.1)
}
=
7.9\% .
\label{eq:imlp_average}
\end{equation}

The mass-splitting dependence of this improvement has a direct physical origin. The oscillation phase,

\begin{equation}
\phi_{\rm osc}
=
1.27\,
\frac{
\Delta m_{41}^{2}L
}{
E_{\nu}
},
\label{eq:osc_phase}
\end{equation}

determines the characteristic energy scale of the disappearance signature. At relatively small $\Delta m_{41}^{2}$, an appreciable oscillation phase is reached preferentially toward the lower energy part of the reactor spectrum. In this region the oscillation produces a broad spectral modification and the correlated fast neutron contribution has a comparatively large effect on the candidate spectrum. An energy-only likelihood is therefore more susceptible to degeneracy between a genuine oscillation-induced spectral change and a change in the correlated background contribution. The MLP score breaks part of this degeneracy because IBD and fast neutron events populate the classifier-score regions very differently. The larger average improvement below $1~\mathrm{eV}^{2}$ therefore reflects the greater role played by background discrimination in this part of the parameter space rather than an enhancement of the oscillation probability itself.

Around $\Delta m_{41}^{2}\sim1$-$2~\mathrm{eV}^{2}$, including much of the RAA-motivated region, the relative 13~m/18~m prompt energy spectra themselves provide a stronger oscillation discriminator. The energy-only likelihood consequently retains a larger fraction of the available oscillation information, and the incremental gain obtained from the classifier score is smaller. At still larger mass splittings, the oscillation phase varies rapidly across both the reactor spectrum and the finite baseline distribution. Detector energy resolution, energy binning, and baseline averaging progressively wash out this fine spectral structure. Event topology can improve the background constraint, but it cannot recover oscillation information that has been physically averaged by the source and detector response. The local structures in $\mathcal{I}_{\rm MLP}(\Delta m_{41}^{2})$ should therefore be interpreted as the combined consequence of the $L/E_{\nu}$ oscillation phase, detector response, and fast neutron spectral composition rather than as additional oscillation resonances.

The statistical origin of the improvement can also be illustrated schematically. Ignoring nuisance correlations for this argument, the information on an oscillation parameter $\vartheta$ in the energy-only analysis scales approximately as

\begin{equation}
\mathcal{F}_{E}
\sim
\sum_{d,i}
\frac{
\left(
\partial S_{di}/\partial\vartheta
\right)^2
}{
S_{di}+B_{di}
},
\label{eq:fisher_energy}
\end{equation}

whereas retaining the classifier-score dimension gives

\begin{equation}
\mathcal{F}_{E,P}
\sim
\sum_{d,i,j}
\frac{
\left(
\partial S_{dij}/\partial\vartheta
\right)^2
}{
S_{dij}+B_{dij}
}.
\label{eq:fisher_2d}
\end{equation}

Because the MLP-score intervals have strongly different signal to background compositions, summing over $j$ before performing the fit removes information that remains available to the two-dimensional likelihood. High-score bins contain a larger fraction of the oscillation-sensitive IBD population, whereas low-score bins constrain the fast neutron normalization and spectral contribution.

This also explains the dependence of the sensitivity improvement on the assumed fast neutron contamination. At a fixed IBD exposure, changing the pre-MLP composition from S/B=1:2 to S/B=1:4 increases the logarithmically averaged improvement from $7.9\%$ to $9.2\%$. The energy-only likelihood then becomes increasingly affected by background fluctuations and by the degeneracy between the fast neutron contribution and an oscillation-induced spectral modification. The topology dimension becomes correspondingly more informative because it supplies internal background control regions. This increase does not indicate a change in the intrinsic MLP classification performance, it reflects the increasing statistical value of an independent signal-background discriminator as the candidate population becomes more background dominated.

The energy dependence of the classifier response remains important for the physical interpretation. For the diagnostic hard requirement $\mathrm{P_{IBD}>}0.40$, the pre-selection IBD spectrum can be reconstructed as

\begin{equation}
\widehat{S}_{di}^{\,\rm corr}
=
\frac{
N_{di}^{\rm pass}
-
\widehat{B}_{di}^{\rm FN,pass}
}{
\epsilon_{di}^{\rm MLP}
},
\label{eq:sterile_corrected}
\end{equation}

where $\epsilon_{di}^{MLP}$ is the energy-dependent IBD acceptance. The spectral-closure study of Sec.~\ref{sec:likelihood} demonstrates that the underlying prompt energy distribution can be recovered after residual background subtraction and efficiency correction when the corresponding statistical and efficiency uncertainties are retained. This provides the direct detector-level connection to the light sterile neutrino analysis. The sensitivity contours of Fig.~\ref{fig:sterile_sensitivity}, however, are not obtained from the hard-selected and corrected sample, they are derived from the full $\mathrm{E_{prompt}\times P_{IBD}}$ likelihood, so that low-score events are retained as background control information rather than discarded.

The projected sensitivity remains conditional on the assumed fast neutron normalization and spectral shape, the detector-response model, and the modeled classifier-score response. This qualification is particularly important because sterile-neutrino oscillations themselves appear as spectral distortions, and an incorrectly modeled correlated background can therefore become partially degenerate with an oscillation hypothesis. A data analysis will require the fast neutron normalization and spectral shape to be constrained using reactor-OFF data and dedicated control samples, together with experimental validation of the energy- and score-dependent classifier response.

The likelihood framework developed here therefore provides a prospective analysis strategy for the existing two-baseline ISMRAN data set and is projected to yield a stronger constraint on the RAA-motivated parameter space than an energy-only analysis. The present contours remain sensitivity projections rather than measured exclusions, but they demonstrate that the same segmented event topology developed for fast neutron suppression can be incorporated directly into the oscillation likelihood and can improve the robustness of a sterile-neutrino search in the presence of correlated backgrounds.

\section{Summary and outlook}

A topology-based multilayer perceptron (MLP) has been developed for the identification of IBD events in the segmented ISMRAN very short baseline neutrino detector, with particular emphasis on the rejection of correlated fast neutron backgrounds. The discrimination exploits the different prompt event topologies produced by positron electromagnetic energy deposition and proton recoils in a segmented scintillator detector array. Three reconstructed interpretable observables describing the energy sharing and energy concentration among detector bars are used as inputs to the classifier.

The dependence of the classifier response on the assumed fast neutron energy spectrum was examined using an independent AmBe application sample, while retaining the same trained network, feature standardization, and fixed threshold. The IBD acceptance in this sample is $76.83\%$, while the fast neutron rejection is $89.71\%$. The difference from the uniformly generated $2$-$12~\mathrm{MeV}$ training population demonstrates that the quantitative background acceptance is spectrum dependent and should not be regarded as a universal classifier property. For a controlled pre-MLP $S/B=1:1$ composition, the fixed selection increases the expected $S/B$ from $1.00$ to $7.47$ and the IBD fraction from $50.0\%$ to $88.19\%$. The total-variation distance of the mixed candidate spectrum from the pre-MLP IBD reference decreases from $0.343$ to $0.055$, corresponding to an $84.0\%$ reduction of the net candidate spectrum deformation. However, the selected signal-only spectrum gives $\mathrm{D_{TV}^{IBD,sel}}=0.123$, demonstrating that the MLP acceptance itself is energy dependent. The particularly small total variation distance of the selected mixed candidate sample therefore results from the combined effects of signal acceptance and residual fast neutron contamination and should not be interpreted as intrinsic preservation of the selected IBD spectral shape.

The energy dependent signal acceptance is propagated explicitly in a detector-level likelihood together with the residual fast neutron component and finite MC template uncertainties. For the nominal pre-MLP $S/B=1:2$ pseudo-experiment, the fixed classifier changes the expected composition to $S/B=3.724$ with an IBD fraction of $78.83\%$. The likelihood recovers the injected signal strength $\mathrm{\alpha_{inj}}=1$ as \[\hat{\alpha}=0.993^{+0.053}_{-0.048},\] with an empirical parametric bootstrap goodness-of-fit probability of $\mathrm{p_{emp}=0.434}$. The extracted pre-MLP spectrum gives the diagnostic closure result $\mathrm{\chi^2_{stat}/N_{dof}}=25.95/34$ and an RMS bin-by-bin deviation of $16.0\%$. Under the benchmark nuisance assumptions adopted for this controlled study, the full relative 68\% uncertainty on $\alpha$ is $5.102\%$ and is dominated by the assumed $5\%$ correlated MLP-efficiency scale. When the pre-MLP fast neutron contribution is doubled to $S/B=1:4$, the post-selection ratio remains $1.862$ and the fit gives $\hat{\alpha}=0.987^{+0.053}_{-0.048}$ with $\mathrm{p_{emp}}=0.613$. The fixed nuisance ensemble exhibits a $-1.02\%$ mean signal-strength bias in this stress configuration, indicating a modest percent level degradation.

Motivated by sterile-neutrino oscillations as a possible explanation of the RAA, the expected active-sterile neutrino oscillation sensitivity is evaluated for the experimentally relevant ISMRAN baselines of 13~m and 18~m, for two years of exposure at each position. The present analysis extends the two baselines oscillation strategy to the 13~m/18~m ISMRAN data-taking configuration and investigates the additional sensitivity obtained when the event-topology information learned by the MLP is incorporated directly into the oscillation likelihood, rather than being used only as a hard event-selection criterion. For the nominal pre-MLP $S/B=1:2$ benchmark, the joint $\mathrm{E_{prompt}\times P_{IBD}}$ likelihood moves the expected 95\% C.L. sensitivity toward the ideal signal-only limit relative to an energy-only analysis with fast neutron contamination, with the largest gain occurring for $\Delta m_{41}^{2}\lesssim 1~\mathrm{eV}^{2}$. In this region, the oscillation-induced prompt energy distortion competes more strongly with the correlated fast neutron contribution, so that the additional event-topology information provides an independent handle for constraining the background. At larger mass splittings, the reconstructed prompt energy spectrum itself carries most of the available oscillation information and the incremental benefit of the classifier score becomes smaller. Over $0.1\leq\Delta m_{41}^{2}\leq1~\mathrm{eV}^{2}$, the resulting $\mathrm{E_{prompt}\times P_{IBD}}$ analysis yields an average $\sim 7.9\%$ stronger 95\% C.L. constraint on $\sin^{2}2\theta_{14}$ than the corresponding prompt energy-only analysis. This demonstrates that the classifier information can contribute directly to the sterile-neutrino physics reach of ISMRAN, rather than serving only as an event-rejection criterion.

The present study therefore establishes a direct connection between segmentation-based event topology and reactor ${\overline{\ensuremath{\nu}}}_{e}$ signal extraction. The MLP suppresses correlated fast neutron backgrounds, while the resulting energy-dependent IBD response can be modeled explicitly and propagated consistently within a spectral likelihood. The analysis further demonstrates that the continuous classifier score can be retained as an additional likelihood dimension, thereby using fast neutron-rich regions to constrain the correlated background without discarding the oscillation information carried by the full prompt-event sample. The residual fast neutron acceptance nevertheless remains dependent on the incident neutron spectrum. Application to the accumulated reactor data will therefore require the fast neutron normalization and spectral shape to be constrained using reactor-OFF or dedicated background samples, together with experimental validation of the energy- and score-dependent IBD response. With these constraints, topology-assisted likelihood analyses can provide an additional handle on correlated backgrounds in segmented reactor ${\overline{\ensuremath{\nu}}}_{e}$ measurements, with applications to reactor monitoring, spectral measurements, and searches for physics beyond the Standard Model.

\section*{Acknowledgements}
The author thanks Dr. Pawan Kumar Netrakanti for valuable comments on the manuscript.

\bibliographystyle{JHEP}
\bibliography{biblio}

@article{STEREO2023Final,
  author = {Almaz{\'a}n, H. and others},
  collaboration = {STEREO},
  title = {{STEREO} neutrino spectrum of $^{235}$U fission rejects sterile neutrino hypothesis},
  journal = {Nature},
  volume = {613},
  pages = {257-261},
  year = {2023},
  doi = {10.1038/s41586-022-05568-2},
  eprint = {2210.07664},
  archivePrefix = {arXiv}
}

@article{kamland,
    author        = {Abe, S. and others},
    collaboration = {KamLAND},
    title         = {Precision Measurement of Neutrino Oscillation Parameters with {KamLAND}},
    journal       = {Phys. Rev. Lett.},
    volume        = {100},
    number        = {22},
    pages         = {221803},
    year          = {2008},
    eprint         = {0801.4589},
    archivePrefix = {arXiv},
    primaryClass  = {hep-ex},
    doi           = {10.1103/PhysRevLett.100.221803}
}

@article{dayabay,
  author = {An, F. P. and others},
  collaboration = {Daya Bay},
  title = {Measurement of the Reactor Antineutrino Flux and Spectrum at Daya Bay},
  journal = {Phys. Rev. Lett.},
  volume = {116},
  pages = {061801},
  year = {2016},
  doi = {10.1103/PhysRevLett.116.061801}
}

@article{doublechooz,
  author = {Abe, Y. and others},
  collaboration = {Double Chooz},
  title = {Indication of Reactor $\bar{\nu}_e$ Disappearance in the Double Chooz Experiment},
  journal = {Phys. Rev. Lett.},
  volume = {108},
  pages = {131801},
  year = {2012},
  doi = {10.1103/PhysRevLett.108.131801}
}

@article{reno,
  author = {Ahn, J. K. and others},
  collaboration = {RENO},
  title = {Observation of Reactor Electron Antineutrino Disappearance in the RENO Experiment},
  journal = {Phys. Rev. Lett.},
  volume = {108},
  pages = {191802},
  year = {2012},
  doi = {10.1103/PhysRevLett.108.191802}
}

@article{prospect,
  author = {Ashenfelter, J. and others},
  collaboration = {PROSPECT},
  title = {The PROSPECT Reactor Antineutrino Experiment},
  journal = {Nucl. Instrum. Meth. A},
  volume = {922},
  pages = {287-309},
  year = {2019},
  doi = {10.1016/j.nima.2018.12.079}
}

@article{stereo,
  author        = {Almaz{\'a}n, H. and others},
  collaboration = {STEREO},
  title         = {Improved Sterile Neutrino Constraints from the {STEREO} Experiment with 179 Days of Reactor-On Data},
  journal       = {Phys. Rev. D},
  volume        = {102},
  pages         = {052002},
  year          = {2020},
  doi           = {10.1103/PhysRevD.102.052002}
}

@article{danss,
  author = {Alekseev, I. and others},
  collaboration = {DANSS},
  title = {Search for Sterile Neutrinos at the DANSS Experiment},
  journal = {Phys. Lett. B},
  volume = {787},
  pages = {56-63},
  year = {2018},
  doi = {10.1016/j.physletb.2018.10.038}
}

@article{dhruva,
  author = {Agarwal, S. K. and Karhadkar, C. G. and Zope, A. K. and Singh, Kanchhi},
  title = {{Dhruva}: Main design features, operational experience and utilization},
  journal = {Nucl. Eng. Des.},
  volume = {236},
  pages = {747-757},
  year = {2006},
  doi = {10.1016/j.nucengdes.2005.09.020}
}

@article{ismran_muon_roni,
  author = {Dey, R. and Netrakanti, P. K. and Mishra, D. K. and Behera, S. P. and Sehgal, R. and Jha, V. and Pant, L. M.},
  title = {Measurement of cosmic muon-induced neutron background with ISMRAN detector in a non-reactor environment},
  journal = {Astroparticle Physics},
  volume = {169},
  pages = {103101},
  year = {2025},
  doi = {10.1016/j.astropartphys.2025.103101}
}

@article{DD_roni,
  author = {Dey, R. and Netrakanti, P. K. and Mishra, D. K. and Behera, S. P. and Mulmule, D. and Patel, T. and Sarkar, P. S. and Jha, V. and Pant, L. M.},
  title = {Characterization of plastic scintillator bars using fast neutrons from {D-D} and {D-T} reactions},
  journal = {JINST},
  volume = {16},
  pages = {P08029},
  year = {2021},
  doi = {10.1088/1748-0221/16/08/P08029},
  eprint = {2110.08299},
  archivePrefix = {arXiv}
}

@article{juno_ml,
  author  = {Gavrikov, A. and Cerrone, V. and Serafini, A. and others},
  title   = {Interpretable machine learning approach for electron antineutrino selection in a large liquid scintillator detector},
  journal = {Phys. Lett. B},
  volume  = {860},
  pages   = {139141},
  year    = {2025},
  doi     = {10.1016/j.physletb.2024.139141}
}

@article{ismran_hql2023,
    author  = {Netrakanti, P. K. and others},
    title   = {Status of {ISMRAN} experiment},
    journal = {PoS},
    volume  = {HQL2023},
    pages   = {026},
    year    = {2024},
    doi     = {10.22323/1.462.0026}
}

@article{ismran_jinst,
  author  = {Mulmule, D. and Netrakanti, P. K. and Pant, L. M. and Nayak, B. K.},
  title   = {Machine learning technique to improve anti-neutrino detection efficiency for the {ISMRAN} experiment},
  journal = {JINST},
  volume  = {15},
  number  = {04},
  pages   = {P04021},
  year    = {2020},
  doi     = {10.1088/1748-0221/15/04/P04021}
}

@article{ismran_dhruv,
  author = {Mulmule, D. and Behera, S. P. and Netrakanti, P. K. and Mishra, D. K. and Kashyap, V. K. S. and Jha, V. and Pant, L. M. and Nayak, B. K. and Saxena, A.},
  title = {A plastic scintillator array for reactor based anti-neutrino studies},
  journal = {Nucl. Instrum. Methods Phys. Res. A},
  volume = {911},
  pages = {104-114},
  year = {2018},
  doi = {10.1016/j.nima.2018.10.026}
}

@article{Mueller,
    author         = "Mueller, Th. A. and others",
    title          = "{Improved predictions of reactor antineutrino spectra}",
    journal        = "Phys. Rev. C",
    volume         = "83",
    pages          = "054615",
    year           = "2011",
    doi            = "10.1103/PhysRevC.83.054615",
    eprint         = "1101.2663",
    archivePrefix  = "arXiv",
    primaryClass   = "hep-ex"
}

@article{Huber,
  author = {Huber, P.},
  title = {Determination of antineutrino spectra from nuclear reactors},
  journal = {Phys. Rev. C},
  volume = {84},
  pages = {024617},
  year = {2011},
  doi = {10.1103/PhysRevC.84.024617},
  note = {Erratum: Phys. Rev. C 85 (2012) 029901},
  eprint = {1106.0687},
  archivePrefix = {arXiv}
}

@article{Vogel,
    author  = "Vogel, P.",
    title   = "{Analysis of the antineutrino capture on protons}",
    journal = "Phys. Rev. D",
    volume  = "29",
    pages   = "1918",
    year    = "1984",
    doi     = "10.1103/PhysRevD.29.1918"
}

@article{ismran_prototype,
  author = {Netrakanti, P. K. and Mulmule, D. and Mishra, D. K. and
            Behera, S. P. and Dey, R. and Sehgal, R. and
            Sinha, S. K. and Jha, V. and Pant, L. M.},
  title = {Measurements using a prototype array of plastic scintillator bars for reactor based electron anti-neutrino detection},
  journal = {Nucl. Instrum. Methods Phys. Res. A},
  volume = {1024},
  pages = {166126},
  year = {2022},
  doi = {10.1016/j.nima.2021.166126}
}

@article{dicebox,
  author  = {F. Be{\v c}v{\'a}{\v r}},
  title   = {Simulation of gamma cascades in complex nuclei with emphasis on assessment of uncertainties of cascade-related quantities},
  journal = {Nucl. Instrum. Methods Phys. Res. A},
  volume  = {417},
  number  = {2-3},
  pages   = {434-449},
  year    = {1998},
  doi     = {10.1016/S0168-9002(98)00787-6}
}

@article{Geant4,
    author  = "Agostinelli, S. and others",
    title   = "{Geant4-a simulation toolkit}",
    journal = "Nucl. Instrum. Meth. A",
    volume  = "506",
    pages   = "250-303",
    year    = "2003",
    doi     = "10.1016/S0168-9002(03)01368-8"
}

@article{ismran_roni_nima,
  author = {Dey, R. and Netrakanti, P. K. and Mishra, D. K. and
            Behera, S. P. and Sehgal, R. and Jha, V. and
            Pant, L. M. and others},
  collaboration = {ISMRAN},
  title = {Evaluation of the response of plastic scintillator bars and measurement of neutron capture time in non-reactor environment for the ISMRAN experiment},
  journal = {Nucl. Instrum. Methods Phys. Res. A},
  volume = {1042},
  pages = {167415},
  year = {2022},
  doi = {10.1016/j.nima.2022.167415}
}

@book{knoll,
  author = {Knoll, Glenn F.},
  title = {Radiation Detection and Measurement},
  publisher = {John Wiley \& Sons},
  address = {Hoboken, NJ},
  edition = {4th},
  year = {2010},
  isbn = {978-0-470-13148-0}
}

@article{ambe,
    author  = {Liu, Cuihong and Pan, Xingyu and Ma, Hao and Zeng, Zhi and Li, Junli and Zhang, Hui},
    title   = {A study on neutron energy spectrum estimation by {LaBr$_3$:Ce} detector},
    journal = {Journal of Radioanalytical and Nuclear Chemistry},
    volume  = {320},
    number  = {3},
    pages   = {859-864},
    year    = {2019},
    doi     = {10.1007/s10967-019-06549-y}
}

@article{BarlowBeeston,
    author  = {Barlow, R. J. and Beeston, C.},
    title   = {Fitting using finite Monte Carlo samples},
    journal = {Comput. Phys. Commun.},
    volume  = {77},
    number  = {2},
    pages   = {219-228},
    year    = {1993},
    doi     = {10.1016/0010-4655(93)90005-W}
}

@article{Mention2011RAA,
    author        = {Mention, G. and Fechner, M. and Lasserre, Th. and
                     Mueller, Th. A. and Lhuillier, D. and Cribier, M. and
                     Letourneau, A.},
    title         = {The Reactor Antineutrino Anomaly},
    journal       = {Phys. Rev. D},
    volume        = {83},
    pages         = {073006},
    year          = {2011},
    doi           = {10.1103/PhysRevD.83.073006},
    eprint        = {1101.2755},
    archivePrefix = {arXiv},
    primaryClass  = {hep-ex}
}

@article{Agostini2020SterileStatistics,
  author = {Agostini, Matteo and Neumair, Birgit},
  title = {Statistical methods applied to the search of sterile neutrinos},
  journal = {Eur. Phys. J. C},
  volume = {80},
  pages = {750},
  year = {2020},
  doi = {10.1140/epjc/s10052-020-8279-x},
  eprint = {1906.11854},
  archivePrefix = {arXiv}
}

@article{Cowan2011Asimov,
  author = {Cowan, Glen and Cranmer, Kyle and Gross, Eilam and Vitells, Ofer},
  title = {Asymptotic formulae for likelihood-based tests of new physics},
  journal = {Eur. Phys. J. C},
  volume = {71},
  pages = {1554},
  year = {2011},
  doi = {10.1140/epjc/s10052-011-1554-0},
  note = {Erratum: Eur. Phys. J. C 73 (2013) 2501},
  eprint = {1007.1727},
  archivePrefix = {arXiv}
}

@article{Behera2020Sterile,
    author        = {Behera, S. P. and Mishra, D. K. and Pant, L. M.},
    title         = {Active-sterile neutrino mixing constraints using reactor antineutrinos with the {ISMRAN} setup},
    journal       = {Phys. Rev. D},
    volume        = {102},
    number        = {1},
    pages         = {013002},
    year          = {2020},
    doi           = {10.1103/PhysRevD.102.013002},
    eprint        = {2007.00392},
    archivePrefix = {arXiv},
    primaryClass  = {hep-ex}
}

\end{document}